\documentstyle[10pt,epsf,epsfig,hangcaption,xspace,amssymb,amsfonts,amsmath,amsthm,cite,
dp_delphititle,lineno]{dp_delphi}
\makeindex
\def\DpPaperGroup{EP}
\def\DpPaperRef{2002-078}
\def\DpDate{4 October 2002}
\def\DpAuthors{DELPHI Collaboration}
\def\DpSubmit{(Accepted by Eur. Phys. J. C)}
\def\DpTitle{Search for  $\mbox{B}^0_s-\overline{\mbox{B}^0_s}$ 
oscillations and a measurement of $\mbox{B}^0_d-\overline{\mbox{B}^0_d}$ 
oscillations using events with an inclusively reconstructed vertex }
\def\DpComment{ }
\def\DpEMail{ }

\newcommand{\dmq}{\Delta m_{q}}
\newcommand{\dmd}{\Delta m_d}
\newcommand{\dms}{\Delta m_s}

\begin{document}
\makeatletter
\newcount\@tempcntc
\def\@citex[#1]#2{\if@filesw\immediate\write\@auxout{\string\citation{#2}}\fi
  \@tempcnta\z@\@tempcntb\m@ne\def\@citea{}\@cite{\@for\@citeb:=#2\do
    {\@ifundefined
       {b@\@citeb}{\@citeo\@tempcntb\m@ne\@citea\def\@citea{,}{\bf ?}\@warning
       {Citation `\@citeb' on page \thepage \space undefined}}%
    {\setbox\z@\hbox{\global\@tempcntc0\csname b@\@citeb\endcsname\relax}%
     \ifnum\@tempcntc=\z@ \@citeo\@tempcntb\m@ne
       \@citea\def\@citea{,}\hbox{\csname b@\@citeb\endcsname}%
     \else
      \advance\@tempcntb\@ne
      \ifnum\@tempcntb=\@tempcntc
      \else\advance\@tempcntb\m@ne\@citeo
      \@tempcnta\@tempcntc\@tempcntb\@tempcntc\fi\fi}}\@citeo}{#1}}
\def\@citeo{\ifnum\@tempcnta>\@tempcntb\else\@citea\def\@citea{,}%
  \ifnum\@tempcnta=\@tempcntb\the\@tempcnta\else
   {\advance\@tempcnta\@ne\ifnum\@tempcnta=\@tempcntb \else \def\@citea{--}\fi
    \advance\@tempcnta\m@ne\the\@tempcnta\@citea\the\@tempcntb}\fi\fi}
 
\makeatother
\begin{titlepage}
\pagenumbering{roman}
\CERNpreprint{\DpPaperGroup}{\DpPaperRef} 
\date{{\small\DpDate}} 
\title{\DpTitle} 
\address{\DpAuthors} 
\begin{shortabs} 
\noindent
 Neutral B meson oscillations in the $\mbox{B}^0_s-\overline{\mbox{B}^0_s}$
 and $\mbox{B}^0_d-\overline{\mbox{B}^0_d}$
 systems  were studied using a sample of about 4.0  million
hadronic Z decays recorded by the DELPHI detector between 1992 and 2000.
 Events with a high transverse momentum lepton were removed and
 a sample of 770 k events with an inclusively reconstructed vertex 
 was selected. 

 The mass difference between the two physical states in the $\mbox{B}^0_d-\overline{\mbox{B}^0_d}$
system 
was measured to be: 

\begin{center}
 $\Delta m_d = (0.531 \pm 0.025 (stat.) \pm 0.007 (syst.)) \rm ps^{-1}$. 
\end{center}
The following limit on the width difference of these states was also obtained:
\begin{center}
 $|\Delta\Gamma_{\rm B_d}| / \Gamma_{\rm B_d} < 0.18$ at 95\% CL. 
\end{center}

As no evidence for  $\mbox{B}^0_s-\overline{\mbox{B}^0_s}$
oscillations was found, a limit on the 
mass difference of the two physical states was given:

\begin{center}
 $\Delta m_s > 5.0 \: \rm ps^{-1}$ at 95 \% CL. 
\end{center}
The corresponding sensitivity of this analysis is equal to 6.6 ps$^{-1}$.
\end{shortabs}
\vfill
\begin{center}
\DpSubmit \ \\ 
\DpComment \ \\
\DpEMail \ \\
\end{center}
\vfill
\clearpage
\headsep 10.0pt
\addtolength{\textheight}{10mm}
\addtolength{\footskip}{-5mm}
\begingroup
%
\newcommand{\DpName}[2]{\hbox{#1$^{\ref{#2}}$},\hfill}
\newcommand{\DpNameTwo}[3]{\hbox{#1$^{\ref{#2},\ref{#3}}$},\hfill}
\newcommand{\DpNameThree}[4]{\hbox{#1$^{\ref{#2},\ref{#3},\ref{#4}}$},\hfill}
\newskip\Bigfill \Bigfill = 0pt plus 1000fill
\newcommand{\DpNameLast}[2]{\hbox{#1$^{\ref{#2}}$}\hspace{\Bigfill}}
\small
\noindent
\DpName{J.Abdallah}{LPNHE}
\DpName{P.Abreu}{LIP}
\DpName{W.Adam}{VIENNA}
\DpName{P.Adzic}{DEMOKRITOS}
\DpName{T.Albrecht}{KARLSRUHE}
\DpName{T.Alderweireld}{AIM}
\DpName{R.Alemany-Fernandez}{CERN}
\DpName{T.Allmendinger}{KARLSRUHE}
\DpName{P.P.Allport}{LIVERPOOL}
\DpName{U.Amaldi}{MILANO2}
% ...

\normalsize
\endgroup
\titlefoot{Department of Physics and Astronomy, Iowa State
     University, Ames IA 50011-3160, USA
    \label{AMES}}
\titlefoot{Physics Department, Universiteit Antwerpen,
     Universiteitsplein 1, B-2610 Antwerpen, Belgium \\
     \indent~~and IIHE, ULB-VUB,
     Pleinlaan 2, B-1050 Brussels, Belgium \\
     \indent~~and Facult\'e des Sciences,
     Univ. de l'Etat Mons, Av. Maistriau 19, B-7000 Mons, Belgium
    \label{AIM}}
\titlefoot{Physics Laboratory, University of Athens, Solonos Str.
     104, GR-10680 Athens, Greece
    \label{ATHENS}}
\titlefoot{Department of Physics, University of Bergen,
     All\'egaten 55, NO-5007 Bergen, Norway
    \label{BERGEN}}
\titlefoot{Dipartimento di Fisica, Universit\`a di Bologna and INFN,
     Via Irnerio 46, IT-40126 Bologna, Italy
    \label{BOLOGNA}}
\titlefoot{Centro Brasileiro de Pesquisas F\'{\i}sicas, rua Xavier Sigaud 150,
     BR-22290 Rio de Janeiro, Brazil \\
     \indent~~and Depto. de F\'{\i}sica, Pont. Univ. Cat\'olica,
     C.P. 38071 BR-22453 Rio de Janeiro, Brazil \\
     \indent~~and Inst. de F\'{\i}sica, Univ. Estadual do Rio de Janeiro,
     rua S\~{a}o Francisco Xavier 524, Rio de Janeiro, Brazil
    \label{BRASIL}}
\titlefoot{Coll\`ege de France, Lab. de Physique Corpusculaire, IN2P3-CNRS,
     FR-75231 Paris Cedex 05, France
    \label{CDF}}
\titlefoot{CERN, CH-1211 Geneva 23, Switzerland
    \label{CERN}}
\titlefoot{Institut de Recherches Subatomiques, IN2P3 - CNRS/ULP - BP20,
     FR-67037 Strasbourg Cedex, France
    \label{CRN}}
\titlefoot{Now at DESY-Zeuthen, Platanenallee 6, D-15735 Zeuthen, Germany
    \label{DESY}}
\titlefoot{Institute of Nuclear Physics, N.C.S.R. Demokritos,
     P.O. Box 60228, GR-15310 Athens, Greece
    \label{DEMOKRITOS}}
\titlefoot{FZU, Inst. of Phys. of the C.A.S. High Energy Physics Division,
     Na Slovance 2, CZ-180 40, Praha 8, Czech Republic
    \label{FZU}}
\titlefoot{Dipartimento di Fisica, Universit\`a di Genova and INFN,
     Via Dodecaneso 33, IT-16146 Genova, Italy
    \label{GENOVA}}
\titlefoot{Institut des Sciences Nucl\'eaires, IN2P3-CNRS, Universit\'e
     de Grenoble 1, FR-38026 Grenoble Cedex, France
    \label{GRENOBLE}}
\titlefoot{Helsinki Institute of Physics, HIP,
     P.O. Box 9, FI-00014 Helsinki, Finland
    \label{HELSINKI}}
\titlefoot{Joint Institute for Nuclear Research, Dubna, Head Post
     Office, P.O. Box 79, RU-101 000 Moscow, Russian Federation
    \label{JINR}}
\titlefoot{Institut f\"ur Experimentelle Kernphysik,
     Universit\"at Karlsruhe, Postfach 6980, DE-76128 Karlsruhe,
     Germany
    \label{KARLSRUHE}}
\titlefoot{Institute of Nuclear Physics,Ul. Kawiory 26a,
     PL-30055 Krakow, Poland
    \label{KRAKOW1}}
\titlefoot{Faculty of Physics and Nuclear Techniques, University of Mining
     and Metallurgy, PL-30055 Krakow, Poland
    \label{KRAKOW2}}
\titlefoot{Universit\'e de Paris-Sud, Lab. de l'Acc\'el\'erateur
     Lin\'eaire, IN2P3-CNRS, B\^{a}t. 200, FR-91405 Orsay Cedex, France
    \label{LAL}}
\titlefoot{School of Physics and Chemistry, University of Lancaster,
     Lancaster LA1 4YB, UK
    \label{LANCASTER}}
\titlefoot{LIP, IST, FCUL - Av. Elias Garcia, 14-$1^{o}$,
     PT-1000 Lisboa Codex, Portugal
    \label{LIP}}
\titlefoot{Department of Physics, University of Liverpool, P.O.
     Box 147, Liverpool L69 3BX, UK
    \label{LIVERPOOL}}
\titlefoot{LPNHE, IN2P3-CNRS, Univ.~Paris VI et VII, Tour 33 (RdC),
     4 place Jussieu, FR-75252 Paris Cedex 05, France
    \label{LPNHE}}
\titlefoot{Department of Physics, University of Lund,
     S\"olvegatan 14, SE-223 63 Lund, Sweden
    \label{LUND}}
\titlefoot{Universit\'e Claude Bernard de Lyon, IPNL, IN2P3-CNRS,
     FR-69622 Villeurbanne Cedex, France
    \label{LYON}}
\titlefoot{Dipartimento di Fisica, Universit\`a di Milano and INFN-MILANO,
     Via Celoria 16, IT-20133 Milan, Italy
    \label{MILANO}}
\titlefoot{Dipartimento di Fisica, Univ. di Milano-Bicocca and
     INFN-MILANO, Piazza della Scienza 2, IT-20126 Milan, Italy
    \label{MILANO2}}
\titlefoot{IPNP of MFF, Charles Univ., Areal MFF,
     V Holesovickach 2, CZ-180 00, Praha 8, Czech Republic
    \label{NC}}
\titlefoot{NIKHEF, Postbus 41882, NL-1009 DB
     Amsterdam, The Netherlands
    \label{NIKHEF}}
\titlefoot{National Technical University, Physics Department,
     Zografou Campus, GR-15773 Athens, Greece
    \label{NTU-ATHENS}}
\titlefoot{Physics Department, University of Oslo, Blindern,
     NO-0316 Oslo, Norway
    \label{OSLO}}
\titlefoot{Dpto. Fisica, Univ. Oviedo, Avda. Calvo Sotelo
     s/n, ES-33007 Oviedo, Spain
    \label{OVIEDO}}
\titlefoot{Department of Physics, University of Oxford,
     Keble Road, Oxford OX1 3RH, UK
    \label{OXFORD}}
\titlefoot{Dipartimento di Fisica, Universit\`a di Padova and
     INFN, Via Marzolo 8, IT-35131 Padua, Italy
    \label{PADOVA}}
\titlefoot{Rutherford Appleton Laboratory, Chilton, Didcot
     OX11 OQX, UK
    \label{RAL}}
\titlefoot{Dipartimento di Fisica, Universit\`a di Roma II and
     INFN, Tor Vergata, IT-00173 Rome, Italy
    \label{ROMA2}}
\titlefoot{Dipartimento di Fisica, Universit\`a di Roma III and
     INFN, Via della Vasca Navale 84, IT-00146 Rome, Italy
    \label{ROMA3}}
\titlefoot{DAPNIA/Service de Physique des Particules,
     CEA-Saclay, FR-91191 Gif-sur-Yvette Cedex, France
    \label{SACLAY}}
\titlefoot{Instituto de Fisica de Cantabria (CSIC-UC), Avda.
     los Castros s/n, ES-39006 Santander, Spain
    \label{SANTANDER}}
\titlefoot{Inst. for High Energy Physics, Serpukov
     P.O. Box 35, Protvino, (Moscow Region), Russian Federation
    \label{SERPUKHOV}}
\titlefoot{J. Stefan Institute, Jamova 39, SI-1000 Ljubljana, Slovenia
     and Laboratory for Astroparticle Physics,\\
     \indent~~Nova Gorica Polytechnic, Kostanjeviska 16a, SI-5000 Nova Gorica, Slovenia, \\
     \indent~~and Department of Physics, University of Ljubljana,
     SI-1000 Ljubljana, Slovenia
    \label{SLOVENIJA}}
\titlefoot{Fysikum, Stockholm University,
     Box 6730, SE-113 85 Stockholm, Sweden
    \label{STOCKHOLM}}
\titlefoot{Dipartimento di Fisica Sperimentale, Universit\`a di
     Torino and INFN, Via P. Giuria 1, IT-10125 Turin, Italy
    \label{TORINO}}
\titlefoot{INFN,Sezione di Torino, and Dipartimento di Fisica Teorica,
     Universit\`a di Torino, Via P. Giuria 1,\\
     \indent~~IT-10125 Turin, Italy
    \label{TORINOTH}}
\titlefoot{Dipartimento di Fisica, Universit\`a di Trieste and
     INFN, Via A. Valerio 2, IT-34127 Trieste, Italy \\
     \indent~~and Istituto di Fisica, Universit\`a di Udine,
     IT-33100 Udine, Italy
    \label{TU}}
\titlefoot{Univ. Federal do Rio de Janeiro, C.P. 68528
     Cidade Univ., Ilha do Fund\~ao
     BR-21945-970 Rio de Janeiro, Brazil
    \label{UFRJ}}
\titlefoot{Department of Radiation Sciences, University of
     Uppsala, P.O. Box 535, SE-751 21 Uppsala, Sweden
    \label{UPPSALA}}
\titlefoot{IFIC, Valencia-CSIC, and D.F.A.M.N., U. de Valencia,
     Avda. Dr. Moliner 50, ES-46100 Burjassot (Valencia), Spain
    \label{VALENCIA}}
\titlefoot{Institut f\"ur Hochenergiephysik, \"Osterr. Akad.
     d. Wissensch., Nikolsdorfergasse 18, AT-1050 Vienna, Austria
    \label{VIENNA}}
\titlefoot{Inst. Nuclear Studies and University of Warsaw, Ul.
     Hoza 69, PL-00681 Warsaw, Poland
    \label{WARSZAWA}}
\titlefoot{Fachbereich Physik, University of Wuppertal, Postfach
     100 127, DE-42097 Wuppertal, Germany \\
\noindent
{$^\dagger$~deceased}
    \label{WUPPERTAL}}

\addtolength{\textheight}{-10mm}
\addtolength{\footskip}{5mm}
\clearpage
\headsep 30.0pt
\end{titlepage}
%
\pagenumbering{arabic} 
\setcounter{footnote}{0} %
\large
\section{Introduction}

 In the Standard Model, $\mbox{B}^0_q-\overline{\mbox{B}^0_q}$ ($q = d,s$) 
mixing  is a 
direct consequence of second order weak interactions. Starting with a
 $\mbox{B}^0_q$ meson produced at time $t$=0, the probability density ${\cal P}$ 
to observe a 
 $\mbox{B}^0_q$ decaying at the proper time $t$ can be written, neglecting 
effects from CP violation:
\begin{center} 
${\cal P}(\mbox{B}^0_q\rightarrow \mbox{B}^0_q)~=~\frac{\Gamma_q}{2} e^{- \Gamma
_{q}t}
[\cosh (\frac{\Delta \Gamma_q}{2} t)~+~\cos (\Delta m_q t) ]$.
\end{center} 
Here $\Gamma_q~=~\frac{\Gamma_q^H~+~\Gamma_q^L}{2}$, $\Delta \Gamma_q~=~\Gamma_q
^H-\Gamma_q^L$,
and $\Delta m_q~=~m^H_q~-~m^L_q$, where $H$ and $L$ denote respectively the heavy 
and light physical states.
The oscillation period gives a direct measurement of the mass difference between
 the two physical states. The
Standard Model predicts that $\Delta \Gamma~\ll~\Delta m$ \cite{ref:Franz}. Neglecting a possible
 difference
between the lifetimes
of the heavy and light mass eigenstates, the above expression simplifies to:
\begin{equation} 
{\cal P}_{\mbox{B}_q^0}^{unmix.}~=~{\cal P}(\mbox{B}_q^0\rightarrow \mbox{B}_q^0)~=
~\frac{1}{2 \tau_{q}} e^{- \frac{t}{\tau_{q}}} [ 1 + \cos ({\Delta m_q t} ) ]
\label{unmix} 
\end{equation}
and similarly:
\begin{equation} 
{\cal P}_{\mbox{B}_q^0}^{mix.}~={\cal P}(\mbox{B}_{q}^{0} \rightarrow \overline{\mbox{B}_{q}^{0}})~ 
=~\frac{1}{2 \tau_{q}} e^{- \frac{t}{\tau_{q}}} [ 1 - \cos ({\Delta m_q t} ) ],
\label{mix} 
\end{equation}
where $\tau_{q}$ is the lifetime of the B$^{0}_{q}$. 
\vskip 0.1cm
\noindent
In the Standard Model, 
the $\mbox{B}^0_q-\overline{\mbox{B}^0_q}$ ($q = \mathrm{d,s}$) 
mass difference $\dmq$ (having kept
only the dominant top quark contribution) can be expressed as follows \cite{ref:Franz}:
\begin{equation}
\dmq= \frac{G_F^2}{6 \pi^2}|V_{tb}|^2|V_{tq}|^2 m_t^2 m_{\rm B_q} f_{\rm B_q}^2 B_{\rm B_q}
\eta_B F(\frac{m_t^2}{m_W^2}).
\label{eq:dmth}
\end{equation}
In this expression $G_F$ is the Fermi coupling constant; $F(x_t)$, with
$x_t=\frac{m_t^2}{m_W^2}$, results from the evaluation of the box diagram and
has a smooth dependence on $x_t$.
$\eta_B$ is a QCD correction factor obtained at next-to-leading order 
in perturbative QCD. 
The dominant uncertainties in Eq.(\ref{eq:dmth}) come from
the evaluation of the B meson decay constant $f_{\rm B_q}$ and of the ``bag'' 
parameter $B_{\rm B_q}$
\cite{ref:bello}.
In terms of the Wolfenstein parametrization \cite{PDG2000}, the two elements of
the $V_{CKM}$ matrix are equal to:
\begin{equation}
 |V_{td}| = A \lambda^3 \sqrt{( 1- \rho )^2 +  \eta^2}  ~~~~~ ;~~~~~        |V_{
ts}| = A \lambda^2,
\end{equation}
neglecting terms of order $O(\lambda^4)$. At this order $\mid V_{ts} \mid$ is 
independent of $\rho$ and $\eta$
and is equal to $\mid V_{cb} \mid$.
Eq. (\ref{eq:dmth}) 
relates $\dmd$ to $|V_{td}|$. It defines 
a circle in the ${\rho}-{\eta}$ plane.
Nevertheless the precision on $\dmd$
cannot be fully exploited due to the large uncertainty
which originates in the evaluation of the non-perturbative QCD parameters.

The ratio between the Standard Model expectations for $\dmd$ and $\dms$ is given
 by:
\begin{equation}
\frac{\dmd}{\dms}~=~\frac{ m_{\rm B_d} f^2_{\rm B_d} B_{\rm B_d} \eta_{\rm B_d}}
{ m_{\rm B_s} f^2_{\rm B_s} B_{\rm B_s} \eta_{\rm B_s}} \frac{\left | V_{td} \right |^2}{\left |
 V_{ts} \right |^2}.
\label{eq:ratiodms}
\end{equation}
 A measurement or a limit on the ratio $\frac{\dmd}{\dms}$ gives a circular constraint
in the ${\rho}-{\eta}$ plane. This ratio depends only on the ratio of the 
non-perturbative QCD parameters which is expected to be better determined than 
their absolute values which occur in Eq. (\ref{eq:dmth}).
Using constraints on $\rho$ and $\eta$ from existing measurements 
(except those on $\dms$), the distribution for
the expected values of
$\dms$ can be obtained. It has been shown that $\dms$ should lie,
at 95$\%$ C.L., between 9.7 and 23.2  ps$^{-1}$
\cite{ref:bello}.

 Using the DELPHI data, several analyses searching for 
 $\mbox{B}^0_s-\overline{\mbox{B}^0_s}$ oscillations have been performed on selected event samples of  
 exclusively reconstructed $\mbox{B}^0_s$ mesons, $\mbox{D}_s$-lepton pairs, $\mbox{D}_s$-hadron pairs 
and events with a high transverse momentum lepton \cite{ref:bspapers}. 
 In this analysis events with a high transverse momentum lepton have been removed
and the remaining events are used to search for $\mbox{B}^0_s$ oscillations and to measure the 
$\mbox{B}^0_d$ oscillation frequency.
Two analyses will be described: one inclusive vertex analysis based on a probabilistic approach using the data set from 1992 to 2000 and one based on neural networks optimized for high values of $\dms$ using only the 1994 data. 
To avoid overlap with other analyses \cite{ref:bspapers}, events with a high transverse momentum
lepton are removed from the sample. Both analyses reconstruct an inclusive secondary vertex which is used to
estimate the proper time. Events that mix are selected using a tag 
based on several separating variables which are combined using 
 probabilities or neural networks respectively. 
 The neural network analysis will provide a check and a confirmation of the
results and in particular of the sensitivity at high values of $\dms$.

 The inclusive vertex analysis
is presented in section 2, describing the secondary vertex and proper time 
reconstruction, the production and decay tags and the fitting programme. 
The measurement of the $\mbox{B}^0_d -\overline{\mbox{B}^0_d}$  oscillation frequency is described 
in section 2.7 and the results of the search for  
$\mbox{B}^0_s-\overline{\mbox{B}^0_s}$ oscillations are presented  in section 2.8.
In section 3, the neural network analysis is described, while 
the conclusions are presented in section 4. 

 The results presented in this paper will be combined later with other DELPHI and LEP
results. 

\section{Inclusive vertex analysis}

 For a description of the DELPHI detector and of its performance
the reader is referred to \cite{ref:delphi}.
 The analysis described in this paper 
used the precise tracking based on the silicon
microvertex detector to reconstruct the primary and secondary
vertex. To estimate the B momentum and direction,   
the neutral particles detected in the electromagnetic 
and hadronic calorimeter and the reconstructed tracks were used. Muon identification 
was based on the hits in the muon chambers being associated with a track.
Electrons were identified using tracks associated with a shower in the 
electromagnetic calorimeter. The dE/dx energy loss measurement in the 
Time Projection Chamber and the Cherenkov light detected 
in the RICH were used to separate pions (and also electrons or muons) 
from kaons and protons. 

Tracks were selected if they satisfied the following criteria:
  their particle momentum was above 200 MeV/c, their tracklength was at least 30 cm, 
their relative momentum error was less than 130\%, their polar angle was 
between 20$^{\circ}$ and 160$^{\circ}$
and their impact parameter with respect to the primary vertex was less than 4 
cm in the $xy$ plane (perpendicular to the beam) and 10 cm in $z$ (along the 
beam direction). 
Neutral particles had to deposit at least 500 MeV in the calorimeters and 
their polar angle had to lie between 2$^{\circ}$ and 178$^{\circ}$.

To select hadronic events it was required that more than 7 tracks of charged particles were accepted with a total
energy above 15 GeV. 
The thrust direction was determined using charged and neutral particles and its polar angle was 
required to satisfy $|\cos(\theta_{thrust})|<0.8$.   The event was divided 
into hemispheres by a plane perpendicular to the thrust axis. In each hemisphere
the total energy from charged and neutral particles had to be larger than 5 GeV.
In total about 4 million hadronic Z decays were selected from which 3.5 million were taken in the LEP I phase (1992-1995) and 500k were 
collected as calibration data in the LEP II phase (1996-2000).
 
Using tracks with vertex detector information, the primary vertex was fitted 
using the average beamspot as a constraint \cite{ref:borissov}.
For each track the impact parameter with respect to the primary vertex was 
calculated and the lifetime sign determined as explained in the paper quoted above.
The b tagging probability\footnote{$E$ refers to the fact that the 
total 
event was used and the $+$ sign means that the lifetime sign had to be 
positive.} $P_E^{+}$ is a measure of the consistency of these track impact parameters
with the hypothesis that all selected tracks came from the event's production
vertex. Events without long-lived particles should have a uniform distribution
of $P^+_E$, while those containing a b-quark tend to have small values.
 In the 1992 and 1993 data the vertex detector measured only the $R\phi$ (R being defined as $\sqrt{x^2+y^2}$ and $\phi$ the azimuthal angle) 
coordinate, while from 1994 to 2000  the $z$ coordinate was also measured.
In the 1992-1993 data, events were selected if the b tagging variable $P^+_E$
 was less than 0.1, whereas in the 1994-2000 data, the cut could be placed at 0.015.

 Jets were reconstructed using charged and neutral particles by the LUCLUS \cite{ref:jetset}  
jet algorithm with an invariant mass cut DJOIN of 6 GeV/c$^2$.
 Leptons were identified and their transverse
 momentum with respect to the jet axis was determined. 
 Loosely identified muons with momenta above
 3 GeV/c were accepted as well as standard and tightly identified muon with  momenta above 2 GeV/c. The reader is referred to \cite{ref:delphi} for the 
identification criteria. 
 Events with a standard or tightly identified muons with momentum
 above 3 GeV/c and a transverse momentum above 1.2 GeV/c were removed from
the selected event sample.
 This was done  
 to avoid overlap with other analyses that use leptons \cite{ref:bspapers}
with a high transverse momentum.
 For electron identification a neural network was used with a cut value
 that corresponds to 75\% efficiency \cite{ref:delphi}. 
 The electron had to have a momentum above 2 GeV/c. 
 Electrons with a momentum below 3 GeV/c had to pass a cut value that 
corresponds to 65\% efficiency. Again
to avoid overlap with other analyses that use high transverse momentum leptons, events with an electron with momentum above 3 GeV/c and a transverse momentum     
 above 1.2 GeV/c satisfying a cut value that corresponds to 65\% efficiency were removed.
The selected electrons and muons will henceforth be referred to as soft leptons. 
 
 Samples of  hadronic Z decays (4 million events) and of Z bosons decaying
only into $\rm b\bar{\rm b}$ quark pairs (2 million events) were 
simulated using the Monte Carlo generator programme JETSET7.3 \cite{ref:jetset}
with DELPHI tuned JETSET parameters and updated b and c decay tables 
\cite{ref:tuning}. 
The detailed response of the DELPHI detector was simulated \cite{ref:detector}. 

 \subsection{Secondary vertex reconstruction} 

 The secondary vertex reconstruction and proper time determination procedures are identical for
 events with or without a soft lepton.
 First the probability $P_i$ that a charged or a neutral particle 
 comes from the secondary (bottom or charm) vertex was parametrized\footnote{Thus a $P_i$ value of 0.8 means that 80 percent of the selected particles will come from the secondary vertex.}.
 The following information was used for tracks: the lifetime-signed impact 
parameters and their errors (in $R\phi$ and $Rz$), 
the transverse momentum with respect to the jet axis,
the muon and electron identification and the rapidity\footnote{For
 calculating rapidities, charged and neutral particles were assigned the pion 
mass.} with respect to the jet axis.
 For neutral particles the transverse momentum and rapidity  were used. 
 For each of these quantities the probability was parametrized using the
simulation. The total probability was obtained by combining these individual
probabilities assuming they are independent.

 To start the first level secondary vertex fit, tracks were selected with at
least one associated hit in the vertex detector and a probability $P_i$ larger
than 60\%. 
The decay length - i.e. the 3-D decay distance - per track was determined by calculating the point of closest 
approach of the track to the B particle trajectory which  was 
approximated by a track
coming from the primary vertex and having the direction of the reconstructed 
jet.  
 The first level secondary vertex  was fitted using  
 the measured decay lengths per track and their errors, the azimuthal and polar angles of the tracks and the B trajectory.
 The result of this approximate fit was a decay length, its error and a $\chi^2$ of the fit. 
Further, the $\chi^2_t$ contribution of each   
track to the total $\chi^2$ was determined. 
 To remove tracks coming from the primary vertex the following iterative procedure
 was performed: 
 if the secondary vertex was reconstructed with more than two tracks, 
 the track upstream of the vertex (i.e. closer to the fitted primary vertex) 
 with the largest $\chi^2_t$ 
 was removed if its $\chi^2_t$ was larger than 4.
 Secondly, tracks were removed that did not combine with any of the other tracks.
 To achieve this, all two track combinations were made and the number of good 
matches  was counted.
 A good match was defined as a 
 two track vertex that was within 2 standard deviations of the 
 fitted secondary vertex.
 For each track, the fraction $f_{good}$ of good matches to the total number 
of combinations was determined.
 The track with the smallest $f_{good}$ value was removed if its value was 
below 20\%, and then the first level vertex fit was redone.
The procedure ends when no track could be removed by the listed criteria.
 
 At the end of this procedure a full vertex fit was performed using
 the measured track parameters and the corresponding covariance matrices. 
 To the list of tracks selected for the fit, the B-track with its 
 covariance matrix  was added as a constraint.
 For each track 
 the impact parameter and its error
 with respect to the fitted secondary vertex  were calculated. 
 The global $\chi^2$ of the fit was defined as the 
 sum of the squares of the track impact parameters divided by corresponding 
 uncertainties (in $R\phi$ and $Rz$).  As a result the B decay length and its 
error were obtained.

 The presence of tracks from charm particle decays in the vertex fit has two effects.
 Firstly, the fitted vertex does not coincide with the B vertex,
 but is some average between the B and D vertex positions. Secondly, the $\chi^2$
 of the vertex increases because of the charm decay length.
 It was therefore important to remove as much as possible the decay products of charmed particles from
 the vertex fit. 
 For this purpose the probability that a track came from
 charm was evaluated on the basis of kinematic and vertex information.
 For example, the momentum distribution of particles from charm, in the 
B rest frame, is
 softer than that for particles from B decays. Secondly, a particle from 
charm decay is produced 
 downstream of the fitted vertex, while a particle from a B hadron originated
 upstream of this vertex. 
 Two new vertex fits were performed. In the first, one
 particle that most likely originates from charm was removed. In the second fit,
 the two particles most likely to come from charm were removed. 

 Using the simulation, an estimate was made of the B decay length and of its 
 error,  
 using as an input the fitted decay length, its associated (or raw) error, the $\chi^2$ and the number of fitted tracks. The expected error on the B decay length was parametrized in the same way.
 This was done for the three vertex fits (removing 0, 1 and 2 particles as described in the previous paragraph).
 Removing 1 or 2 particles has the advantage of reducing the  bias
 caused by the presence of particles from charm. On the other hand the 
 resolution is degraded if a track is removed.
 Due to the fact that the $\chi^2$ is sensitive to the presence of
 particles from charm, part of the bias is corrected for in the 
parametrization of the B decay length. 
 Finally, out of the three vertex fit results, the result with the smallest expected error on the B decay length
was chosen. In 51\% of the cases no track was removed, in 36\% one track and 
in 13\% two tracks were removed. 

\begin{figure}[htb]
\vspace*{-0.5cm}
\begin{center}
\epsfysize14.0cm\epsfxsize14.0cm
\epsffile{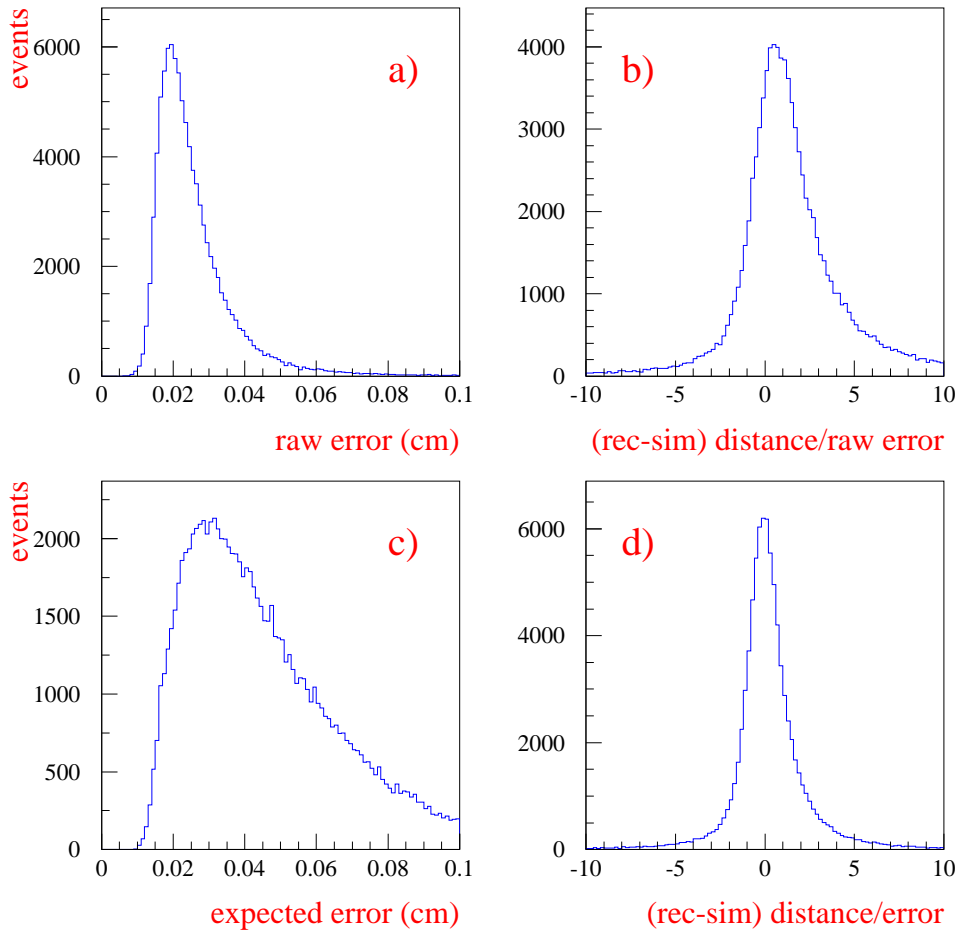}
\caption{Figure a) shows the expected or raw error, Figure b) 
 the reconstructed minus the simulated B decay
distance divided by the raw error for the 1994-1995 simulation.
Figures c) and d) show the expected error and the reconstructed minus the
simulated B decay distance after applying the procedure described in the text.} 
\label{vtx} 
\end{center}
\end{figure}

In Figures \ref{vtx}a and b the raw error as it comes out of the full vertex 
fit and the reconstructed minus the B decay 
distance divided by the raw error are shown for the 1994-1995 simulated events.
The tail due to the presence of charmed particles can be clearly observed.
Figures \ref{vtx}c and d show the expected error and the reconstructed minus the 
simulated B decay distance after applying the correction procedure described 
above. The latter distribution is clearly more Gaussian and its width  
is close to unity.

 \subsection{Proper time reconstruction}

 To determine the proper time, the momentum of the B hadron had to
be measured. An estimate of the energy of the b jet was made, applying 
energy-momentum conservation to the whole event. 
The masses of the jet containing the B hadron
and of the system formed by the remaining charged and neutral particles,
labelled respectively $M_1$ and $M_2$, were measured.
 The b jet energy was obtained as: 
\begin{equation}
E_{jet} = E_{cms}/2-(M_2^2-M_1^2)/(2\:E_{cms}),
\end{equation}
 where $E_{cms}$ is the centre of mass energy. This significantly improved the b jet 
energy resolution. 
 The B energy was determined as: 
\begin{equation}
 E_B=\frac{\sum_i P_i E_i}{\sum_i E_i} E_{jet},
\end{equation}
 where $E_i$ is the energy of the charged  
or neutral particle and $P_i$ is the probability that a particle comes from
the decay of a B hadron (see section 2.1).   

\begin{figure}[htb]
\vspace*{-0.5cm}
\begin{center}
\epsfysize14.0cm\epsfxsize14.0cm
\epsffile{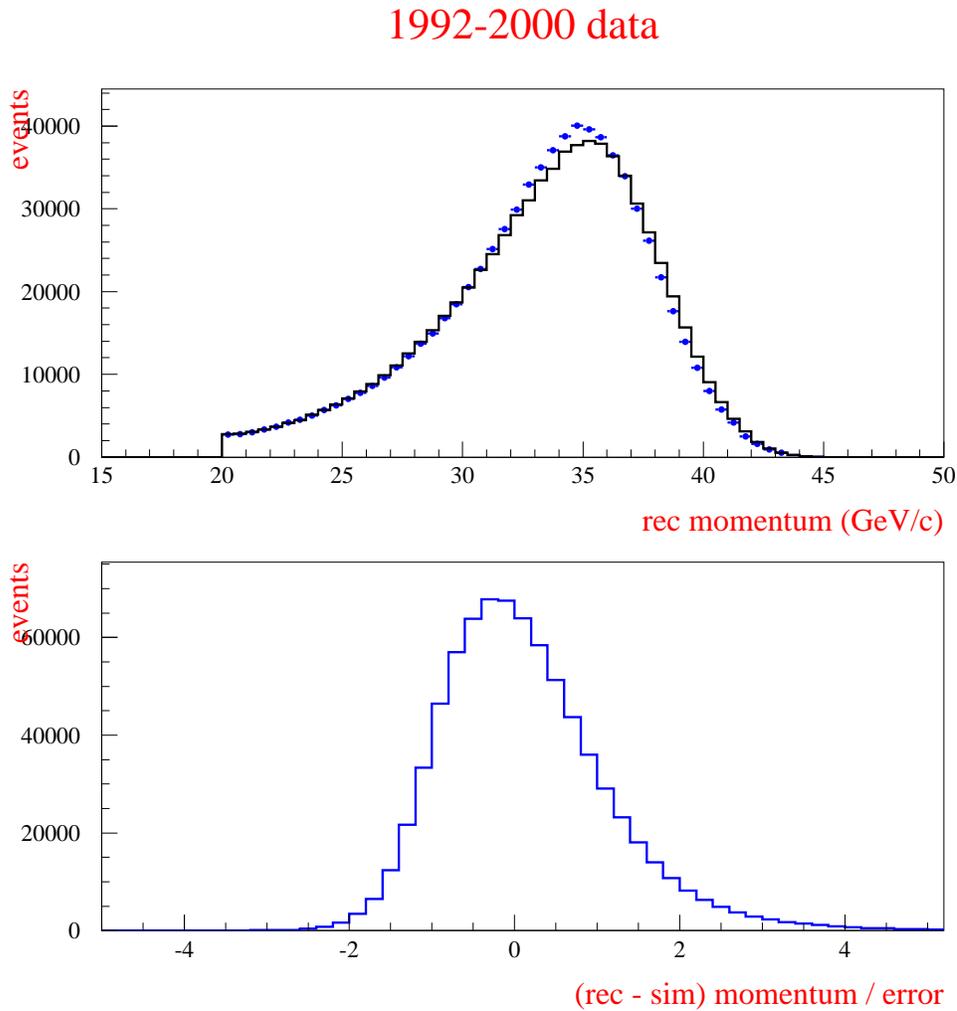}
\caption{Upper diagram: Reconstructed B momentum. The dots correspond to the 
1992 to 2000 data, the solid line to the measured momentum distribution 
obtained from simulated events. Lower diagram: The reconstructed minus the true B 
momentum divided by the expected error for simulated events. }
\label{bmom} 
\end{center}                                                        
\end{figure}

 The momentum of the B hadron was determined from the B energy 
and a small correction typically of order 10\% was applied as a function of the following quantities:
the weighted (with $P_i$) number of charged and neutral particles,
the ratio of the  raw B energy ($\sum_i P_i E_i$) to the jet energy 
$E_{jet}$, the invariant mass $M_1$,   
 the ratio of the charged over the total raw B energy 
 and the number of jets.  The corrections were obtained from the simulation. 
 The reconstructed B momentum is shown in Figure \ref{bmom}.
     
 The expected error was parametrized as a function of the uncorrected B energy 
 and of the jet energy. 
 It lies between 3 and 9 GeV/c and is on the average equal to 5 GeV/c. 
 The reconstructed minus simulated B momentum divided 
 by the expected error for simulated events is shown in Figure \ref{bmom}. 

 The proper time $t$ was calculated using: 
\begin{equation}
\label{for:tau}
  t = \frac{m L}{p}, 
\end{equation}
 where $m$ is the B mass, $L$ the decay length
 and $p$ the estimated B momentum. 
 The expected error $\sigma_t$ on the proper time was estimated using: 
\begin{equation}
 \sigma_t = \sqrt {{\Large\lgroup} \frac{ m \delta L}{p} {\Large \rgroup} ^2  +  {\Large \lgroup} \frac{m L \delta p}{p^2} {\Large \rgroup} ^2}, 
\end{equation} 
 where $\delta L$ is the expected error on the decay length  and $\delta p$ is
 the error on the momentum.
 The data were divided into eight categories according to the 
 value of the proper time resolution. 
 This division was made because most of the sensitivity at high values of
 $\dms$ came from events with the best proper time resolution.
  The cuts are given in Table \ref{categories}.
 To fall into the first category, the expected resolution had to be smaller 
than  0.12+0.07$t$ ps ($t$ in ps units). 
 Events with a resolution worse than 0.35+0.2$t$ ps were rejected.     

\begin{figure}[htb]
\vspace*{-0.5cm}
\begin{center}
\epsfysize14.0cm\epsfxsize14.0cm
\epsffile{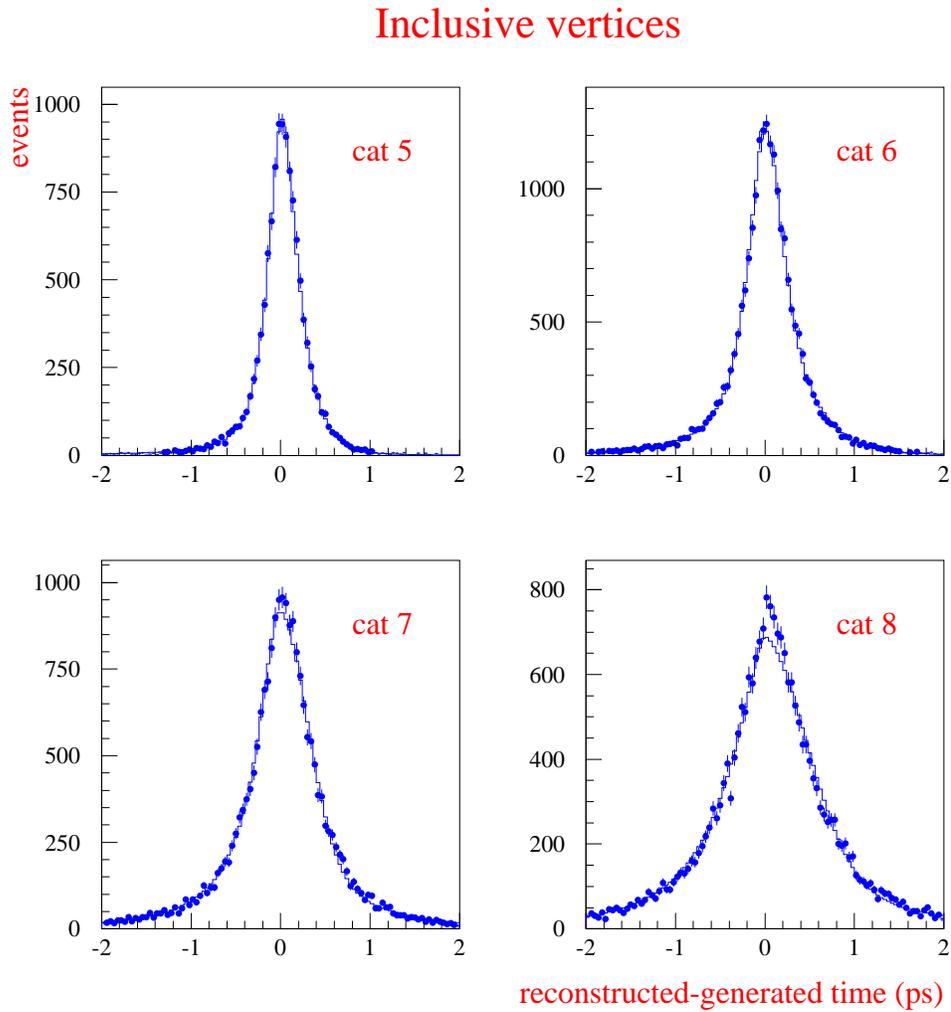}
\caption{Reconstructed minus generated proper time for the inclusive 
B vertex sample corresponding to categories 5 to 8. The dots correspond to the 
simulated data and the histograms to the parametrization of the
resolution function (see section 2.5).}
\label{proptv} 
\end{center}
\end{figure}

 The first four categories refer to events with a soft 
 lepton and the last four to events with only an inclusive vertex. 
 The soft lepton sample consists of 155023 events.
 The latter sample will be referred to as the inclusive vertex sample
 and consists of 614577 events.
 The proper time resolutions for the last four classes are shown in Figure 
 \ref{proptv}.

\begin{figure}[htb]
\vspace*{-0.5cm}
\begin{center}
\epsfysize14.0cm\epsfxsize14.0cm
\epsffile{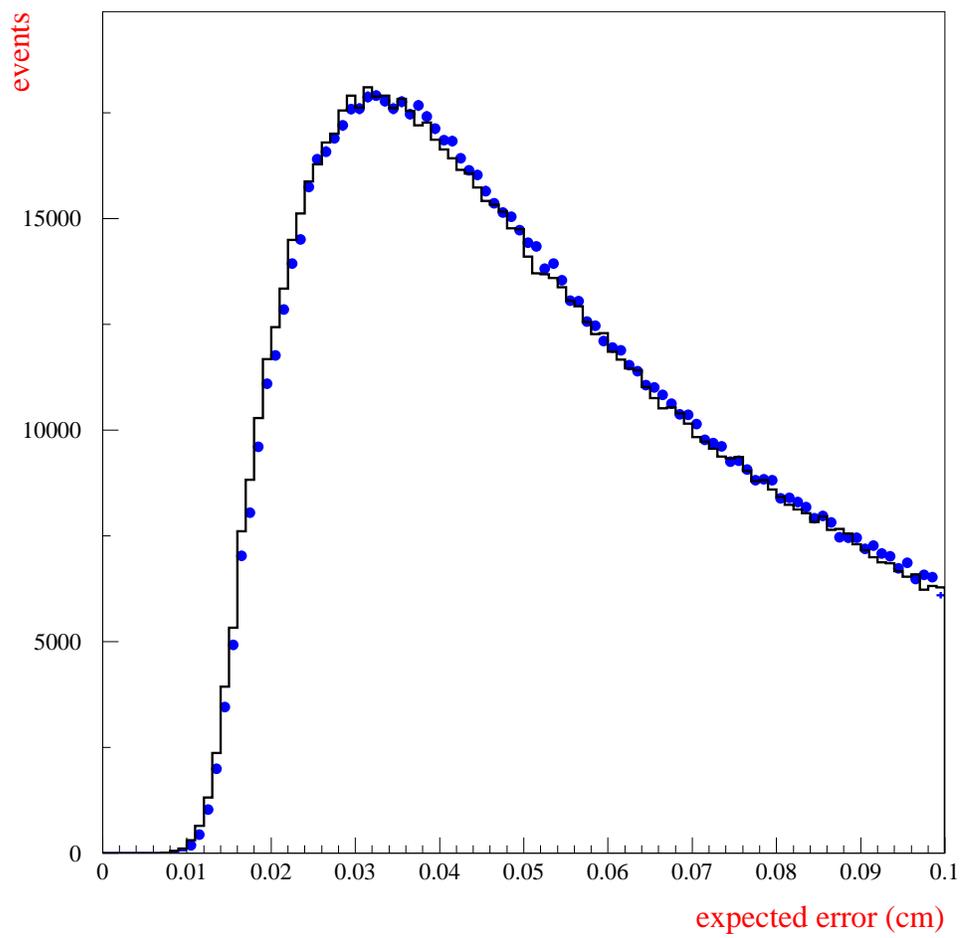}
\caption{The expected error on the decay length for 1992-2000 data 
 (points with error bars)
 and simulation (solid line).}
\label{vtxerror} 
\end{center}                                                        
\end{figure}

 The systematic error on the decay length resolution was estimated to be $\pm$10\%.
 This number was obtained in the following way. First, a comparison
 of data and simulation for the expected decay length error (see Fig. \ref{vtxerror}) showed
 that the data show a discrepancy for a scale error of less than $\pm$5\%. Secondly,
  the description of the impact parameters of the tracks with negative lifetime
  sign allow for a scaling of the associated error of less than $\pm$5\% \cite{ref:borissov}.
  Finally, a study was made of the amplitude error (see section 2.8) as a function of
  $\dms$ comparing data and simulation. The amplitude error increases because
   of the finite proper time resolution. The amplitude error for 
data and simulation  
   are in agreement within $\pm$10\%. 
  This is mainly due to the fact that the numbers of events in each category agree for data and simulation.

  The systematic error on the momentum resolution was estimated to be $\pm$10\%.
  This number was obtained in the following way.  Comparing the observed
  momentum in a hemisphere with the expected momentum in that hemisphere
   - obtained using energy and momentum conservation - for data and simulation,
  it was found that the momentum resolution agreed to better than $\pm$10\%.
  Finally, the study of the amplitude error, mentioned above,
  showed that the amplitude error for data and simulation
  was in agreement within $\pm$10\%.

%
 \begin{table}[htb]
 \begin{center}
 \vspace*{-0cm}
 {\small 
 \begin{tabular}{|c|c|c|c|c|} \hline
   category& 1& 2 & 3 & 4 \\ \hline  
   $\sigma_t (ps)$&  0.12+0.07 $t$ &  0.18+0.08 $t$ &0.25+0.1 $t$ & 0.35+0.2 $t$\\ 
   soft lepton sample& 22740 (5533) & 41597 (10598)  & 42835 (12091) & 47851 (15620) \\ \hline 
  category & 5& 6 & 7 & 8 \\ \hline
   $\sigma_t (ps)$&  0.12+0.07 $t$ &  0.18+0.08 $t$ &0.25+0.1 $t$ & 0.35+0.2 $t$\\ 
  inclusive vertex sample  & 68875 (16476) & 146075 (36633) & 171859 (47702) & 227768  (73809)\\
 \hline
\end{tabular}
}
\caption{Cuts on the resolution $\sigma_t$ and total number of selected events (in parenthesis the number of events corresponding to the 92-93 data) for the different categories.}
\label{categories}
\vspace*{-1cm} 
 \end{center}
 \end{table}

\clearpage 
 \subsection{Production and decay tag} 

 To distinguish between events in which the neutral B meson has mixed 
or not, a production and a decay tag were defined. 
They give,
respectively, the b-flavor content (ie b or $\overline{\rm b}$) of the B hadron at production
and decay times. 
In this analysis both the production and decay tags were optimized for $\mbox{B}^0_s$ mesons. 
In Z decays,  b and $\overline{\rm b}$ quarks are produced back to back, in pairs. 
 The hemisphere $opposite$ to the decaying B can therefore be used to tag the 
flavour at production time. 
 This will be called the opposite side production tag which is obtained from 
a combination of several variables:
 
 $\bullet$ the average charge of a sample of tracks, attached to the
b-jet, and enriched in b-decay products:
 \begin{center}
 $Q_{jet} = \sum q_i p_{iL} / \sum p_{iL}$ with $P_i>0.5$,
 \end{center} 
 
 where $p_{iL}$ is the component of the momentum of the particle along the jet axis direction,
and $P_i$ is its probability that it is a B decay product, as defined at the beginning of section 2.1;   

 $\bullet$  the average charge of a sample of tracks, attached to the
b-jet, and enriched in b-fragmentation products:\\
 \begin{center}
 $Q_{f} = \sum q_i p_{iL} (P_i<0.5) / \sum p_{iL} ({\rm all}\: P_i)$.
 \end{center} 
 Note that the denominator sums over all particles, because the fraction
 of the total longitudinal momentum that is coming from fragmentation particles is
 relevant; 

 $\bullet$ the charge and momentum $p^\star$ of any identified lepton, in the B 
 rest frame, which is determined from the inclusively reconstructed B momentum vector;  

 $\bullet$ the heavy particle charge for an identified kaon or proton
 and its momentum  $p^\star$ in the B rest frame.

\begin{figure}[htb]
\vspace*{-0.5cm}
\begin{center}
\epsfysize14.0cm\epsfxsize14.0cm
\epsffile{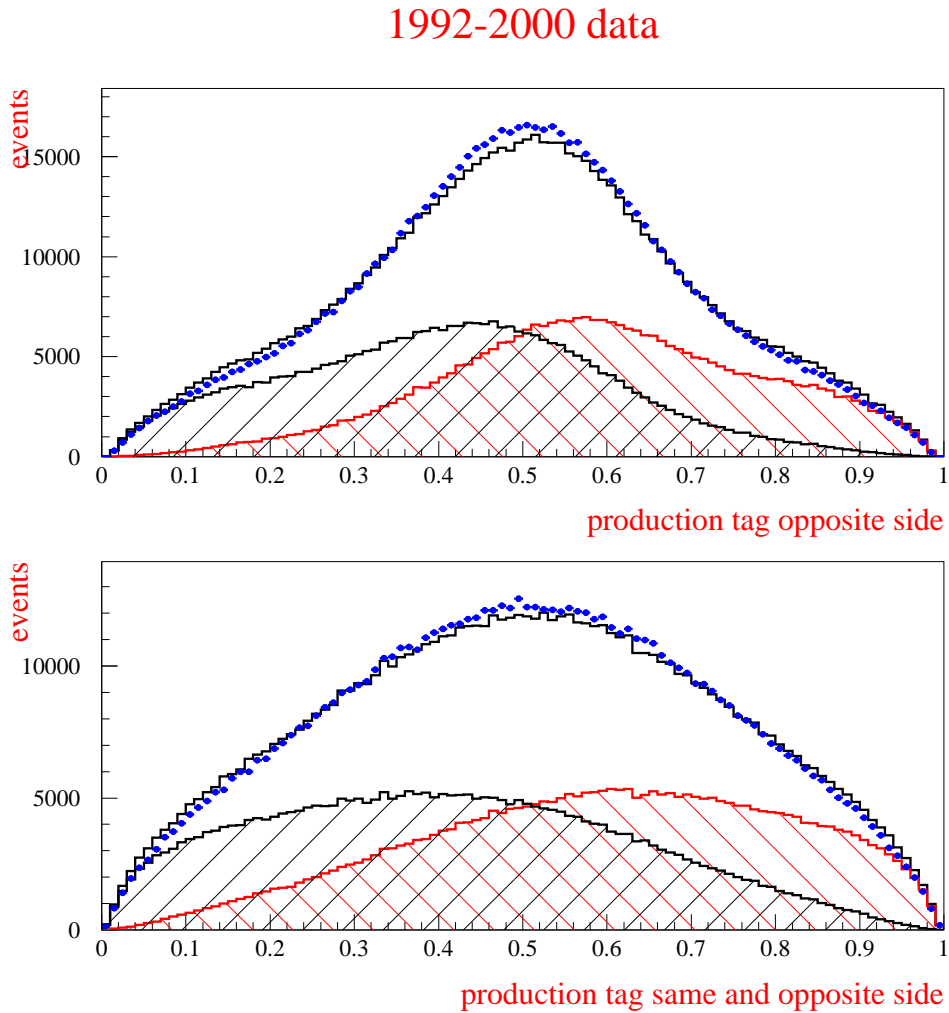}
\caption{Production tag using only information from the opposite side (upper diagram) and the production tag using information from both sides (lower diagram).
 The dots correspond to the 1992 to 2000 data, the solid line to the simulation. The hatched areas correspond to the b (left) and $\bar{\rm b}$ (right) contributions.}
\label{protag}
\end{center}
\end{figure}

 Using simulation, distributions for these variables were obtained for 
 B and $\bar{\rm B}$ hadrons.
 These variables were converted into probabilities $Pb_i$ for the $\bar{\rm b}$  hypothesis,
and then combined to give the opposite side production tag. 
This was done in the 
following way.  For each variable a rejection factor $R_{i}$ is defined as $\frac{Pb_{i}}{1-Pb_i}$
and a combined rejection factor $R$
is obtained by taking the product of the rejections $R_i$.
The combined probability $P$ is then equal to $\frac{R}{1+R}$. 
In Figure \ref{protag} the distribution of the opposite side production tag is shown for 
 1992-2000 data and simulation. 
 The tagging purity is defined as the fraction of correct flavour assignments at 
 100\% efficiency, i.e.  all events were classified if the cut on the combined
 probability was set at 0.5.  A purity equal
to 68\% has been measured on 1992-2000 simulated events.

A {\it same} side production tag is also defined using the fragmentation tracks 
accompanying the decaying B meson.  
Both leading fragmentation pions and kaons are sensitive
to the b or $\bar{\rm b}$ production flavour.
 The following quantity $Q_{same}$ was defined:\\
 \begin{center}
 $Q_{same}= \sum R(p_{iL},h_i) (1-P_i) q_i$, 
 \end{center}
 where $h_i$ is equal to 1 for a heavy (proton, kaon) or to 0 for a light (electron, muon or pion) particle and the sum extends over all tracks.  
 The parametrization of the function $R$  - a polynome as a function of $p_{iL}$ - was obtained using simulated events. 
 The variable $Q_{same}$  was converted into a probability and then combined
 with the opposite side production tag to give the combined production tag $P_{prod}$. 

 In Figure \ref{protag} the distribution of $P_{prod}$  is shown for 
 1992-2000 data and simulation. The uds and charm quark contributions 
are 
small (see Table 2) and are included in the total distribution.
 The tagging purity for  $\mbox{B}^0_s$ mesons  
 is equal to 71\% for the 1992-2000 simulation. 
 As expected, this value is higher than the result, 64\%, obtained using the opposite side 
 production alone \cite{ref:bspapers}.
 The difference between data and simulation for the combined production tag, which is apparent in Fig. \ref{protag} will be taken into
 account by fitting the tagging purity for the data (see 
 section 2.6). 

 The other important variable in the analysis is the decay tag. 
 In the soft lepton sample 
this tag is relatively straightforward using the charge of the lepton. Most of 
the $\mbox{B}-\overline{\mbox{B}}$ separation 
 comes from the momentum $p^\star$ of the lepton in the B rest frame that 
allows the separation of a prompt lepton coming from the B vertex from a 
lepton coming from a charm decay. Other information in the
event (such as, for example, the impact parameter of the lepton 
with respect to the secondary vertex and the isolation of the lepton 
(presence of other tracks from charm decay vertex)) 
helps to improve the 
 $\mbox{B}-\overline{\mbox{B}}$  separation. Finally,
also the decay tag developed for the inclusive vertex sample (discussed below) was added to 
improve the performance slightly.  

 For events with no lepton, obtaining a decay tag is more difficult.
 The following approach was taken. All the charged and neutral particles 
were boosted back in the B reference frame using the estimated B momentum and direction (see section 2.2). The B-thrust axis
was determined in the B reference frame using charged and neutral particles with $P_i$ greater than 0.5. The particles were assigned to the forward
or backward hemisphere. Usually one hemisphere contains
most of the tracks from the B vertex while the other contains
most of the tracks from charm decay. 
 This is called a dipole, as the  $\mbox{\rm B}^0_s$
decays to a $\mbox{\rm D}_s^{-(\star)}$ and a virtual $\mbox{\rm W}^+$ and the 
charge difference between
the two hemispheres is equal to $\pm$ 2.
 Under the hypothesis that the forward (backward) hemisphere contains the 
particles from the charm decay and the backward (forward) hemisphere the particles
from the B vertex, the flavour probability of the decaying $\mbox{B}^0_s$ is evaluated. 
This is achieved by using the 
charge and the momentum $p^{\star}$ in the B rest frame of the heavy 
($p,K$) and light ($e$,$\mu$,$\pi$) particles.
 For these parametrizations, the simulation was used.  
 Then a hemisphere probability is evaluated for the hypothesis that the charmed
particle is in the forward (backward) hemisphere. This probability depends on 
the lifetime-signed impact parameter of the tracks with respect to the secondary vertex,
on their momenta in the B rest frame and on the hemisphere multiplicity.
By combining the hemisphere probability with the flavour probability, 
the decay tag for the inclusive vertex sample was obtained. 
The tag was optimized for $\rm B_{s}$ mesons. 

\begin{figure}[htb]
\vspace*{-0.5cm}
\begin{center}
\epsfysize14.0cm\epsfxsize14.0cm
\epsffile{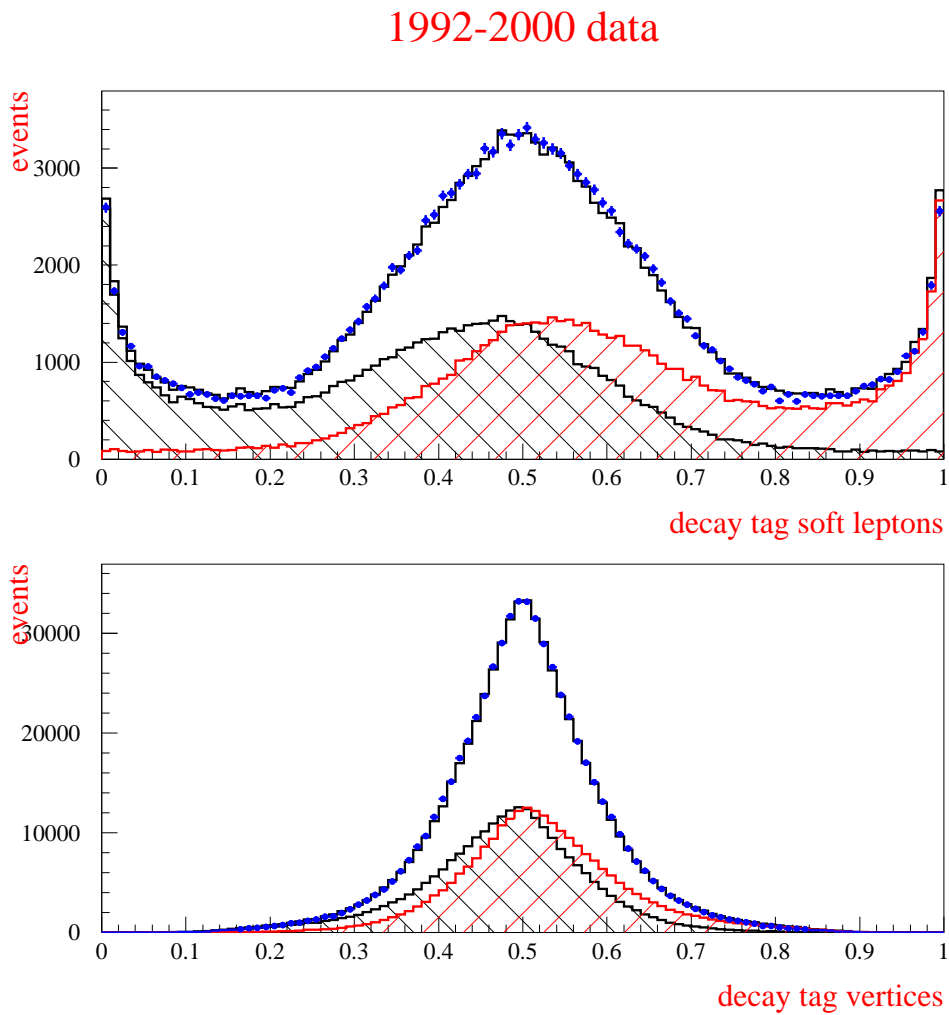}
\caption{Decay tag for the soft lepton sample (upper diagram) and inclusive vertex sample (lower diagram). The 
dots correspond to the 1992 to 2000 data, the solid line to the simulation. 
The hatched areas correspond to the b (right) and $\bar{\rm b}$ (left) contributions at the 
time of decay.}
\label{decaytag} 
\end{center}
\end{figure}

In Figure \ref{decaytag} the performance of the decay tag $P_{decay}$ for the 
soft lepton sample 
is shown for 1992-2000 data and simulation.
The uds and charm quark contributions are
 small (see Table 2) and not shown explicitly.
The tagging purity is 69\% at 100\% efficiency.   
The events with $P_{decay}$ from 0 to 0.02 and 0.98 to 1 are due to prompt B 
decays with a high $p^{\star}$ value.
The performance of the decay tag for the inclusive vertex 
 sample is also shown.
The $\mbox{B}^0_s$ tagging purity is 58\% at 100\% efficiency.   
The difference between data and simulation for the decay tag will be taken into
account by fitting the purity for the data, as is discussed in 
section 2.6.

 \subsection{Sample composition }
\label{sample} 
 For the sample composition the following B-hadron production 
fractions were  assumed \cite{ref:bosc}: 
     $f_{\rm B_s}$ = 0.097 $\pm$ 0.011,
     $f_{\rm B \: baryons}$= 0.103 $\pm$ 0.017,
     $f_{\rm B_u}$ =   
    $f_{\rm B_d}$ = 0.40.  
  For the lifetime of the different B species it was assumed that 
 \cite{ref:bosc}:
     $\tau_{\rm B_u}$ = 1.65 ps, 
     $\tau_{\rm B_d}$ = $\tau_{\rm B^0_s}$ = 1.55 ps and  
     $\tau_{\rm B_{baryons}}$ = 1.20 ps. 

  Using the simulation, the uds and charm backgrounds were extracted.
  The background fractions for the different data sets and vertex 
  categories are listed in Table \ref{bkg}, where $f_{uds}$ is defined as 
  the number of uds events divided by the total number of events in the
  sample.

 \begin{table}[htb]
 \begin{center}
 \vspace*{-0cm}
 \begin{tabular}{|c|c|c|c|c|c|c|c|c|c|c|} \hline
  background & data set & cat 1 & cat 2 & cat 3 & cat 4 & cat 5 & cat 6 & cat 7& cat 8 \\ \hline 
 $f_{uds}$  & 1992-1993  &  .0074 &  .0158 &  .0288 &  .0495 &  .0226 &  .0407 &  .0717 &  .1237 \\
  $f_{uds}$ & 1994-2000  & .0046 &  .0076 &  .0117 &  .0229  & .0138 &  .0199 &  .0329 &  .0588 \\
 $f_{charm}$ & 1992-1993 &  .0202  & .0653 &  .1116 &  .1779 &  .0359 &  .0920  & .1433 &  .1900 \\
  $f_{charm}$ & 1994-2000 & .0356 &  .0673  & .1201 &  .1919 &  .0436 &  .0928 &  .1514 &  .2004 \\
 \hline
\end{tabular}
\caption{The background fractions for the 1992-2000 data sets divided according
to the different vertex categories.}
\label{bkg}
 \end{center}
 \end{table}

  \subsection{Fitting programme}
  \label{ssec:fitting}

  The fitting programme provided an analytic description of the data for the like- and unlike-sign tagged events. It was used to fit the amplitude of 
 $\mbox{B}^0_s-\overline{\mbox{B}^0_s}$ oscillations.
  In the fitting program the 
  time resolution function ${\cal R}(t_{rec}-t_{true},t_{true})$ was parametrized.
  The resolution function gives the probability that, given a certain value 
  for the true proper time $t_{true}$, a proper time value $t_{rec}$ is 
  reconstructed. 
  Two asymmetric Gaussian distributions\footnote{The asymmetric Gaussian has two widths, one 
 for proper time values above the central value, the other for below.} are used 
  to describe the main signal, as well as one asymmetric Gaussian 
to describe  poorly measured events 
  and one Gaussian  to describe the probability
  that the secondary vertex is reconstructed near to the primary vertex.
  The widths of the Gaussian distributions are of the form 
  $\sigma=\sqrt{\sigma_0^2+\sigma_p^2 t_{true}^2}$ with $\sigma_p$ being the relative 
  momentum resolution. The relative normalizations of the Gaussian distributions 
  are left free to vary and parametrized as a constant plus a term proportional to
  $1-e^{-t_{true}/\tau}$, where $\tau$ is the average b lifetime.
  For each vertex category the time resolution function was fitted and 
  the result of the fit is shown in Figure \ref{proptv}. 
 The effect of different parametrisations of the resolution function was found to be neglible with respect to the effect of a change in the proper time resolution (see section 2.2). 
 
  The probability ${\cal P}_b (t_{rec})$ for a B event to be observed at a 
proper time  $t_{rec}$  was written
  as a convolution of an exponential B decay distribution with lifetime $\tau$, 
  an acceptance function $A(t)$ and
  the resolution function: 

 \begin{equation} 
  {\cal P}_b (t_{rec})  = \int_{0}^{\infty} 
    A(t) {\cal R}(t_{rec}-t,t) \frac{e^{-t/\tau}}{\tau} dt.
 \end{equation} 

 The acceptance function was parametrized for the different vertex categories using the simulation. The difference in acceptance for the different B species was found to be negligible. 
 Due to the requirements on the flight distance in the track selection, the 
acceptance is a smooth,  but not flat, function
 of the proper time.  
 The probabilities for uds (${\cal P}_{uds}$) and charm (${\cal P}_c$) events for the different vertex categories are parametrized as a function
 of $t_{rec}$ with exponential functions whose slopes are determined using the simulation. 

 Like- or unlike-sign tagged events are those events for which $P_{comb}$ is respectively larger or smaller than 50\%.
 The combined tagging
 probability $P_{comb}$ is defined as 
 \begin{equation} 
   P_{comb} =  P_{prod}P_{decay} + (1-P_{prod})(1-P_{decay}).
 \label{comb}
 \end{equation} 
 The tagging purity $\epsilon_{\rm B_q}$ is expressed in terms of the combined tagging
 probability $P_{comb}$. For $\mbox{B}^0_s$ events it is given by: 
 \begin{equation} 
  \epsilon_{\rm B_s}= 0.5 + |P_{comb} -0.5| 
 \label{pur}
 \end{equation} 
 The tagging purities for the other B particles and for the charm and light quark
 background events were also expressed as functions of $P_{comb}$ ($P_{prod}$ and
 $P_{decay}$)  
 using the simulation (see section 2.6).

 The total probability to observe a like-sign tagged event at the reconstructed proper time $t_{rec}$ is:

 \begin{eqnarray} 
 {\cal P}^{like}(t_{rec}) = f_b \sum_{q=d,s} f_{\rm B_q} \epsilon_{\rm B_q} {\cal P}^{mix.}_{rec. \rm B_q}(t_{rec}) +
                           \mathnormal f_b \sum_{q=u,d,s,baryons}   f_{\rm B_q} (1- \epsilon_{\rm B_q}) {\cal P}^{unmix.}_{rec. \rm B_q}(t_{rec}) &+ \nonumber \\   
                           f_c (1-\epsilon_c) {\cal P}_c(t_{rec}) + 
                           f_{uds} (1-\epsilon_{uds}) {\cal P}_{uds}(t_{rec})   
 \label{likes}
 \end{eqnarray} 
and correspondingly for an unlike-sign tagged event:
 \begin{eqnarray} 
 {\cal P}^{unlike}(t_{rec}) = f_b \sum_{q=d,s}  f_{\rm B_q} (1-\epsilon_{\rm B_q}) {\cal P}^{mix.}_{rec. \rm B_q}(t_{rec}) +
                      \mathnormal     f_b \sum_{q=u,d,s,baryons}   f_{\rm B_q} \epsilon_{\rm B_q} {\cal P}^{unmix.}_{\rm B_q}(t_{rec}) &+ \nonumber \\   
                           f_c \epsilon_c {\cal P}_c (t_{rec}) +  
                           f_{uds} \epsilon_{uds} {\cal P}_{uds} (t_{rec}).  
 \label{unlikes}
 \end{eqnarray}

 For the mixed $\mbox{B}^0_d$ and $\mbox{B}^0_s$ mesons one has the 
following expression ($q=d,s$):
 \begin{equation} 
   {\cal P}^{mix.}_{rec. \rm B_q} (t_{rec}) = \int_{0}^{\infty} 
  A(t) {\cal R}(t_{rec}-t,t) {\cal P}^{mix.}_{\rm B_q} (t) dt ,
 \end{equation} 
 while for the unmixed case the $\mbox{\rm B}_u$ and B baryons also have to be 
included ($q=u,d,s$,baryons): 
 \begin{equation} 
   {\cal P}^{unmix.}_{rec. \rm B_q} (t_{rec}) = \int_{0}^{\infty}
 A(t) {\cal R}(t_{rec}-t,t) {\cal P}^{unmix.}_{\rm B_q} (t) dt,
 \end{equation} 
 where ${\cal P}^{(un)mix.}_{\rm B_q} (t)$ 
 are defined in Eqs. (\ref{unmix}) and (\ref{mix}). 

  \subsection{Modelling the simulation and data} 
 \label{Modelling}
 The present analysis uses the production and decay probabilities
on an event-by-event basis. The tagging purity 
 $\epsilon_{\rm B_s}$ is calculated from these quantities
 as defined in Eq. (\ref{pur}). 
 The production and decay probabilities - and thus the tagging purities - for 
 the different B species, charm and light quarks are different.
 These differences have to be parametrized in the analytic
 fitting programme.
 For this simulated
events were used. 

\begin{figure}[htb]
\vspace*{-0.5cm}
\begin{center}
\epsfysize14.0cm\epsfxsize14.0cm
\epsffile{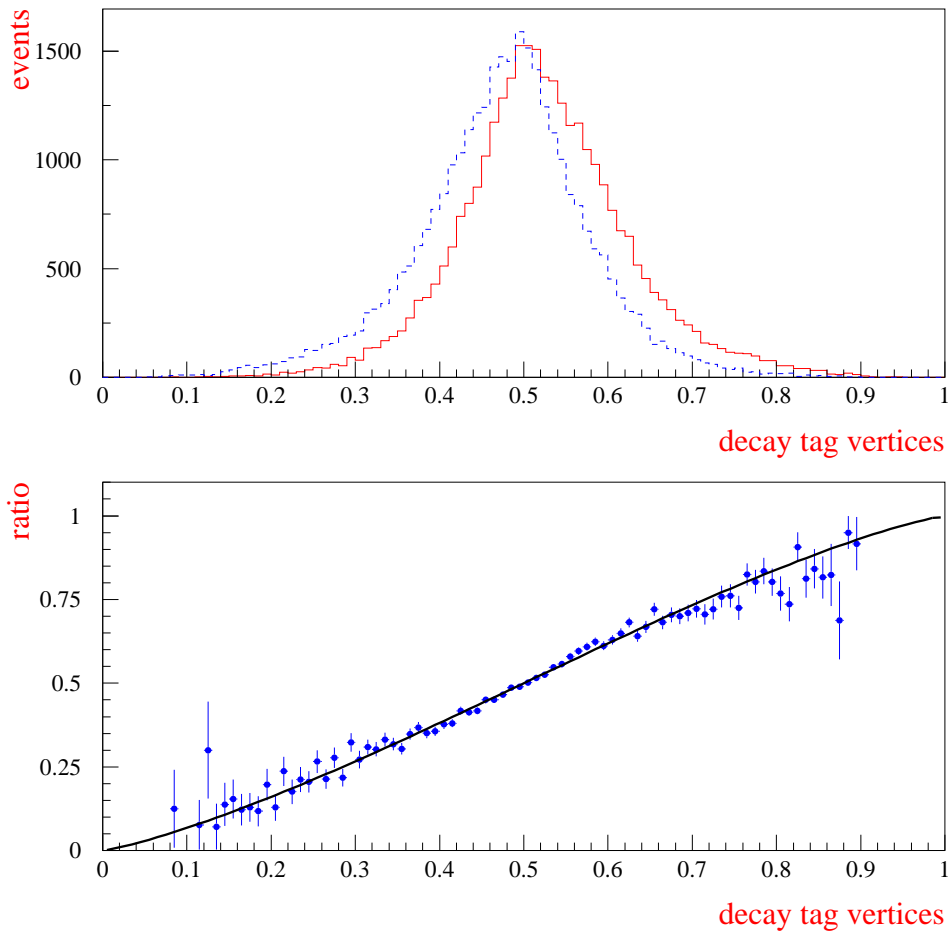}
\caption{Distributions for the decay tag for the inclusive vertex sample for the $\rm B_d$ (solid line) and 
 $\overline{\rm B_d}$ (dashed line) mesons. The bottom plot shows the ratio 
 of the
 two distributions with a fit of Eq.(\ref{eq:slope}) giving the $\alpha$ parameter of 1.15.}
\label{Bdplot} 
\end{center}
\end{figure}

 The new parametrization is obtained by 
 modifying the probability $P$ 
 ($P_{prod}$ or $P_{decay}$). For this purpose a parameter $\alpha$ is introduced
 and the new probability is defined as: 
 \begin{equation} 
   \label{eq:slope}
  P_{new} = R^{\alpha}/(1+R^{\alpha}), 
 \end{equation} 
 where the rejection $R$ is defined as $\frac{P}{1-P}$.
 A parameter value of 1 means that the probability remains unchanged.
It was found out on simulation that this particular parametrisation gives 
an accurate description of the tag performance for neutral B species. 
For example  the $\alpha$ parameter for a $\rm B$ hadron is obtained in
 the following way. Using the simulation the distribution of the probability
 $P$ is plotted separately for $\rm B$ and $\overline{\rm B}$ hadrons. The two distributions are
 divided and fitted to the expression (17) leaving free the $\alpha$ parameter.
 This procedure is illustrated in Fig. \ref{Bdplot} using the decay tag for the inclusive vertex sample 
 for the $\rm B_d$ meson.
 \begin{table}[b]
 \begin{center}
 \vspace*{-0cm}
 \begin{tabular}{|c|c|c|c|} \hline
  Particle   & inclusive vertex sample & soft lepton sample & all events    \\ 
     & decay tag $\alpha_D$ & decay tag $\alpha_D$ & production tag  $\alpha_P$   \\ \hline 
   ${\rm B_s}$ & 1 & 1 & 1 \\
   ${\rm B_d}$ & 1.15 &1 & 1.13 \\
   ${\rm B_u}$ & 0.75 to 1 & 1 & 0.3 to 0.8  \\
   ${\rm B \:baryon}$ &  0.80& 1 & 1.09\\
   uds & 0.20 & 0.20 & 0.80  \\       
   charm & 4.2 (P=0.2-0.8) & 4.2 (P=0.2-0.8) & 0.50 \\
 \hline
\end{tabular}
\caption{The parameters $\alpha$ for the production and decay tag for the different 
particles as obtained from simulation.} 
\label{slopes}
\end{center}
\end{table}

 For leptons, the decay purities for the different B species were studied on simulation and found to be very similar.
 The decay tag parameter for the soft lepton sample was therefore put equal to 1.
 The decay tag parameters for the inclusive vertex sample and soft lepton sample as well as the
 production tag parameter are 
 listed in Table \ref{slopes}. The values are obtained from the simulation.
 For the charm quark a parameter $\alpha_D$ of 4.2 is used if the probability lies between
 0.2 and 0.8, otherwise $\alpha_D$=1. For the soft lepton sample the 
relatively high value of   
$\alpha_D$ of 4.2 
is understandable, because a lepton coming from the charmed particle,
at relatively low $p^{\star}$, will tag correctly the charge of the charmed 
particle.  
 Note that the parameters  $\alpha_D$ and $\alpha_P$ for the different B species 
are quite similar, except for the $\mbox{B}_u$ where $\alpha_D$ varies as a function of $P$ \footnote{The functional form used is $\alpha = \alpha_0 + \alpha_1 \:e ^{2|P-0.5|}$.} between 0.75 and 1. For this reason, tagging purities
for $\mbox{B}_u$ and for the other types of b hadrons have been controlled directly from the 
data, as explained in the following.
 From the new probability $P_{new}^{\rm B_{q},uds,c}$, the combined probability
 $P_{comb}^{\rm B_{q},uds,c}$ is calculated using Eq. (\ref{comb}) and  
 the purity $\epsilon_{\rm {B_q},uds,c}$ is obtained 
 using: 

 \begin{equation} 
  \epsilon_{{\rm B_q,uds,c}}= 0.5 + |P_{comb}^{\rm B_q,uds,c} -0.5|. 
 \end{equation}

 It is important also to have a correct modelling of the tagging purity for the
 data i.e. to have a good description of the like- and unlike-sign tagged events.
 Using the data for each category, a correction 
 factor $C$ to the parameter $\alpha$ is fitted:  

 \begin{equation} 
\label{eq:slopedata}
 \alpha_{D}^{data}=\:C\:\: \alpha_D, 
 \end{equation} 
 where $C$ is determined from the fraction of like-sign tagged events.
 The $C$ factor was determined iteratively and $\dmd$ = 0.531 (see section 2.7) was finally used. 
 For the soft lepton sample, the results are shown in Table \ref{slopel}.
 
 \begin{table}[htb]
 \begin{center}
 \vspace*{-0cm}
 \begin{tabular}{|c|c|c|} \hline
    data set&category & fitted value for $C$  \\ \hline
    1992-1993& 1 &  $0.95 \pm 0.05$\\
           & 2 & $0.81 \pm 0.05$\\
           & 3 & $0.76 \pm 0.05$\\
           & 4 & $0.82 \pm 0.05$\\ \hline 
     1994-2000& 1&  $0.77 \pm 0.04$\\
            & 2&$0.69 \pm 0.04$\\
            & 3&$0.83 \pm 0.04$ \\
            & 4&$0.84 \pm 0.05$ \\
 \hline
\end{tabular}
\caption{The fitted correction factor $C$ for the soft lepton sample.}
\label{slopel}
\end{center}
\end{table}
 The 1992-1993 and 1994-2000 data sets have different
 performance for tracking and lepton identification and therefore
 the fitted $C$ values can be different. 
 The total error on $C$ for the soft lepton sample  
 is better than $\pm$ 5\%. 
 
 The parameter $\alpha$ for the decay tag in the inclusive vertex sample  for a $\mbox{B}_u$ meson 
 is different - it also varies as a function of the tagging probability -from those for the other 
 B particles (see Table \ref{slopes}).  
 By separating the inclusive vertex sample into one enriched  
 and one depleted in $\mbox{B}_u$ particles it was possible
 to determine, from the data, the correction factor $C$ for $\mbox{B}_u$ mesons and for the other
 particles. These samples were obtained by cutting on the secondary vertex charge. 
 Three fits were performed. First it was assumed that the correction factors $C$
 for all types of B particles were identical. Then in a second fit it was assumed that 
 $C$ for the non-$\mbox{B}_u$ particles was equal to 1 and the 
 value of the correction factor $C$ for the $\mbox{B}_u$ particles was fitted. 
 From the $\chi^2$ of the fit it was
 clear that the second fit result was preferred.
 The value of the correction factor $C$ for $\mbox{B}_u$ particles was 
 fixed to the average value between the first and second fit results and 
 a third fit was performed leaving $C$ free for non-$\mbox{B}_u$ particles. 
The results for the final fit are shown in Table \ref{slopev}, where the 
errors quoted in the third column correspond to the statistical errors 
obtained in the first fit. 
 \begin{table}[htb]
 \begin{center}
 \vspace*{-0cm}
 \begin{tabular}{|c|c|c|c|} \hline
    data set&category & correction factor $C$  & correction factor $C$  \\  
    & &  for $\mbox{B}_d$, $\mbox{B}_s$ and $\mbox{B}_{baryon}$ & for $\mbox{B}_u$ mesons   \\ \hline 
    1992-1993 & 5& $0.75 \pm 0.07$ &   0.54 \\
             & 6 & $0.76 \pm 0.06$ &   0.54  \\
             & 7 & $0.72 \pm 0.07$ &   0.40 \\
             & 8 & $0.63 \pm 0.09$ &   0.20 \\ \hline 
    1994-2000&5  & $0.93 \pm 0.05$ &   0.60  \\
             & 6 & $0.94 \pm 0.04$ &   0.60 \\
             & 7 & $0.83 \pm 0.07$ &   0.40 \\
             & 8 & $0.63 \pm 0.09$ &   0.20 \\
 \hline
\end{tabular}
\caption{The fitted correction factors $C$ for the inclusive vertex sample.}
\label{slopev}
\end{center}
\end{table}
 If another procedure was chosen a different $C$ value would have been obtained. 
The largest change in the $C$ value for non-$\mbox{B}_{u}$ particles is quoted as a systematic error and amounts to $\pm$ 15\%. The systematic error  
 is larger than the statistical error. 

 It was found, using the simulation, that the acceptance for the uds and charm
 quarks depends on the tagging purity. 
 The acceptance $A(t)$ for B events also varies slightly as 
 a function of the tagging purity. 
 This was taken into account in the like- and unlike-sign probability distributions.
 A comparison between data and simulation showed a slightly different 
 acceptance function. The acceptance function was corrected to obtain better
 agreement between the data and the parametrisation in the fitting program.
 Note that for $\mbox{B}_d^0-\overline{\mbox{B}_d^0}$ and $\mbox{B}_s^0-\overline{\mbox{B}_s^0}$ oscillations, only the 
fraction of like-sign tag events is relevant and  to first order
the acceptance correction  drops out. 

\begin{figure}[htb]
\vspace*{-0.5cm}
\begin{center}
\epsfysize14.0cm\epsfxsize14.0cm
\epsffile{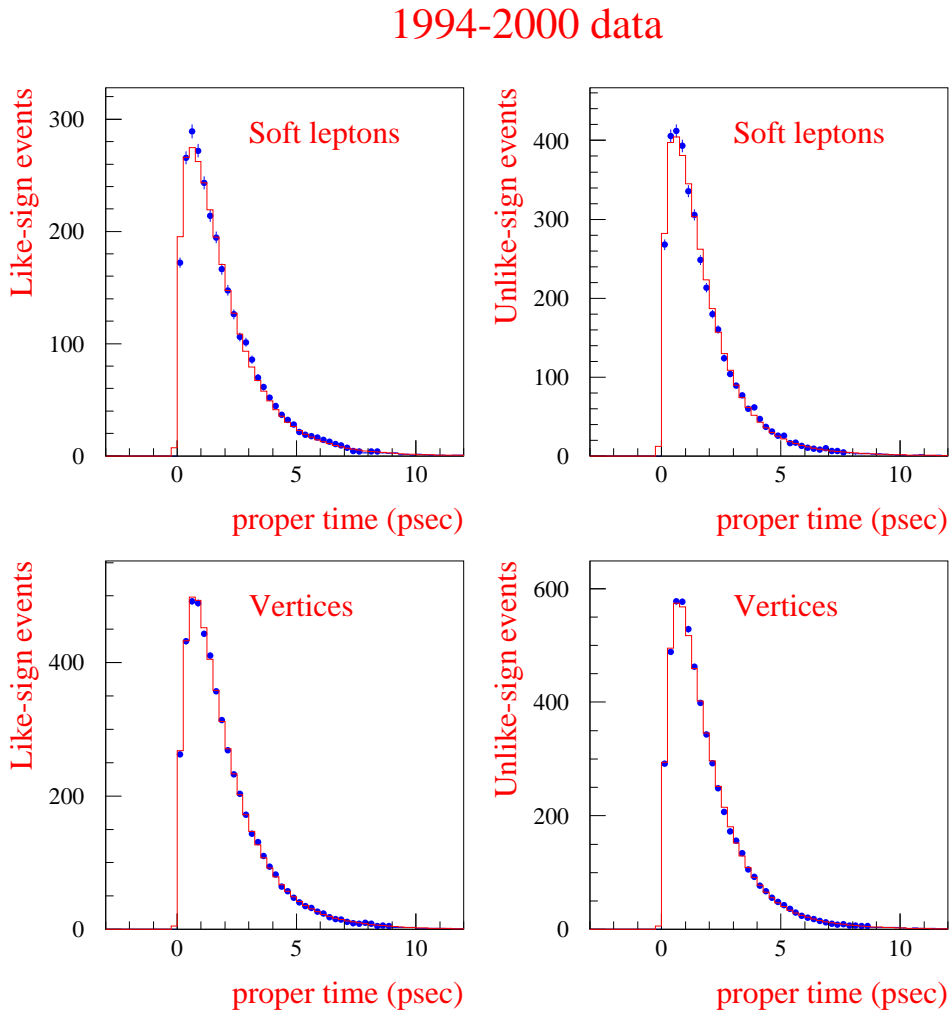}
\caption{Reconstructed proper time distributions for the like- and unlike-sign tagged events corresponding to the soft lepton (upper diagram) and inclusive vertex (lower diagram) samples. The 1994 to 2000 data are shown as dots, the fitted parametrization is shown as a histogram.}
\label{acc94} 
\end{center}
\end{figure}

 In Figure \ref{acc94} the distributions for the like- and 
 unlike-sign tagged events, as  a function of the proper time, 
 corresponding respectively  to the soft lepton sample and to the inclusive vertex sample, 
are shown. In these Figures, the events have 
been 
weighted by $|\epsilon_{\rm B_s}-0.5|$. In this way events with a higher tagging 
purity acquire 
a higher weight. Events with a purity of 0.5 carry no information and 
have a weight equal to zero.  
A good description of the data is obtained.

   
\begin{figure}[htb]
\vspace*{-0.5cm}
\begin{center}
\epsfysize14.0cm\epsfxsize14.0cm
\epsffile{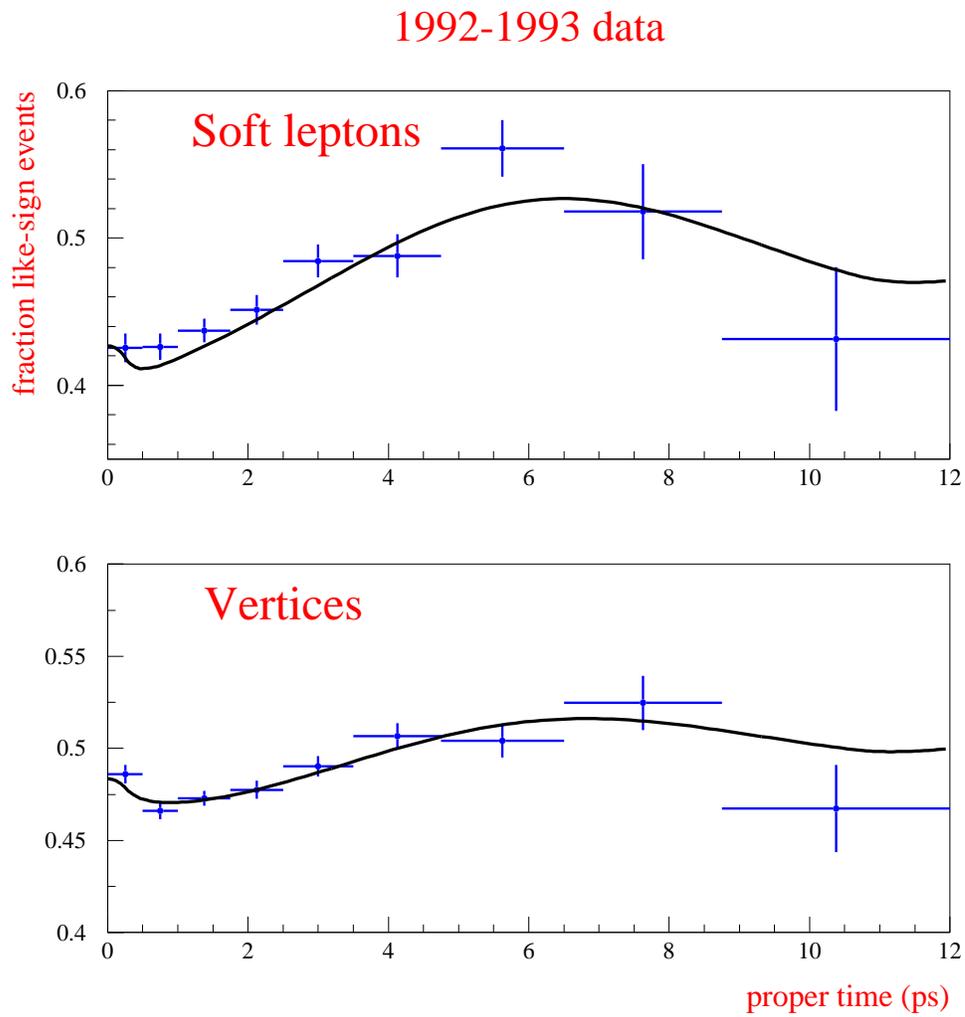}
\caption{Fraction of like-sign tagged events as a function of the reconstructed
proper time for the soft lepton and inclusive vertex samples. 
The 1992 and 1993 data are shown as points with error bars, the parametrization is given as a solid line.}
\label{like9293}
\end{center}
\end{figure}

\begin{figure}[htb]
\vspace*{-0.5cm}
\begin{center}
\epsfysize14.0cm\epsfxsize14.0cm
\epsffile{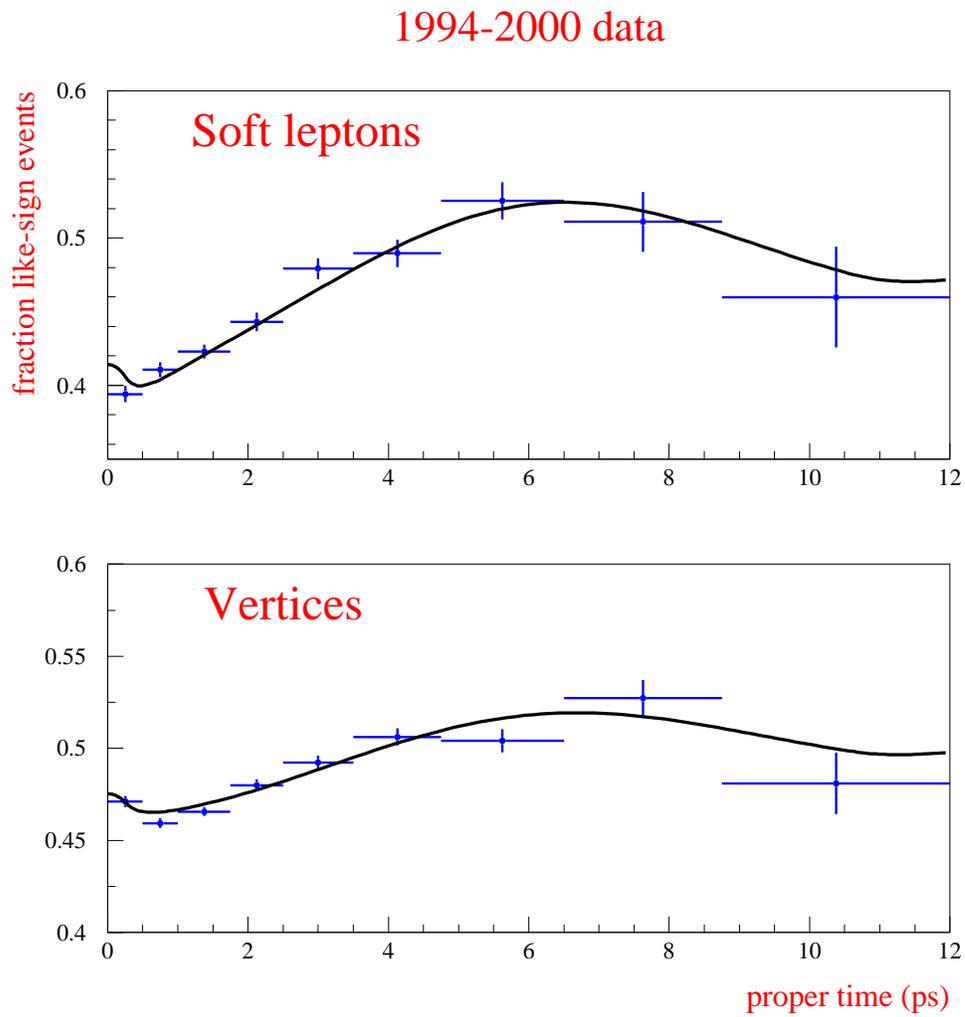}
\caption{Fraction of like-sign tagged events as a function of the 
reconstructed proper time for the soft lepton and inclusive vertex samples. The 1994 to 2000 data are shown as points with error bars, the parametrization is given as a solid line.}
\label{like9495}
\end{center}
\end{figure}

In Figures \ref{like9293} and  \ref{like9495} the fractions of weighted - as described above - like-sign tagged 
events, as a  function 
of the proper time, for the soft lepton sample and inclusive vertex sample, 
are shown for 
the 1992 to 2000 data. In these Figures, values of $\Delta m_d$ of 0.495 ps$^{-1}$ 
 and $\Delta m_s$ of 15 ps$^{-1}$ are used in the parametrizations corresponding to the continuous lines.

\clearpage 
  \subsection{Measurement of  the $\mbox{B}_d$ oscillation frequency } 
 The mass difference between the two physical states in the $\mbox{B}^0_d-\overline{\mbox{B}^0_d}$ system was determined by fitting the fraction
 of weighted like-sign tagged events - shown in Figures \ref{like9293}, \ref{like9495} -  as a function of the reconstructed proper time $t$. 
 The following expression was used for the number of weighted like-sign events: 

 \begin{eqnarray}
\begin{array}{ll}
 
  N_{like}(t) & = A(t)\:  N_{b} \: f_{\rm B_u}\frac{e^{-t/\tau_{\rm B_u}}} {\tau_{\rm B_u}} (1-\epsilon_b)\\  
  & + A(t)\: N_{b} \: f_{\rm B_{baryon}}\frac{e^{-t/\tau_{\rm B_{baryon}}}} {\tau_{\rm B_{baryon}}} (1-\epsilon_b)\\  
 &          + A(t) \: N_{b} \: f_{\rm B_s}\frac{e^{-t/\tau_{\rm B_s}}}{2\tau_{\rm B_s}} \\
       &+ A(t) \: N_{b} \: f_{\rm B_d} \frac{e^{-t/\tau_{\rm B_d}}}{2\tau_{\rm B_d}}  [ (1-2\epsilon_d) \cos(\Delta m_d t) 
           + 1  ] 
\\
& + (1-\epsilon_c) N_c(t) + (1-\epsilon_{uds}) N_{uds}(t).
\end{array}
\label{eq:ar1}
\end{eqnarray}
 The total number of weighted events is equal to:
 \begin{equation} 
  N_{tot}(t) =   \sum A(t) \:N_{b} \: f_{\rm Bi} e^{-t/\tau_{Bi}}/\tau_{Bi} + N_c (t) + N_{uds} (t). 
\label{eq:ar2}
 \end{equation} 
 The event-by-event tagging purity is used as a weight. 
  The values for $f_{\rm Bi}$ and for the B-hadron lifetimes were fixed at the values listed in
  section 2.4. 
$N_b$ is the total number of b quark events.
   The functions $N_c(t)$ and $N_{uds}$ as well as the acceptance $A(t)$ were parametrized using the simulation. 
 The total number of events from charm and uds quarks are obtained by intergrating these functions.
 The tagging purities $\epsilon_c$ and 
 $\epsilon_{uds}$ were taken from the simulation. 

The like-sign tagged fraction $\frac{N_{like}}{N_{tot}}$ was fitted in the range 
from 0.5 to 12 ps using a binned $\chi^2$ fit. First a fit was performed on the simulated data, i.e. 
the parametrization as shown in Figures \ref{like9293} and \ref{like9495}. In this fit 
 $\dmd$, the $\mbox{B}_d$ mass difference, is fixed and  
 $\epsilon_d$, $R_1$, $R_2$ and $a$ are left free, where
  $\epsilon_d$ is the $\mbox{B}_d$ tagging purity and
 the tagging purity for the other b hadrons  $\epsilon_b$ is parametrized as:
  $\epsilon_b =R_1\:e^{at} +R_2 \:e^{-t}$.
  The parameter $R_2$ takes into account the slight dependence
  of the tagging purity as a function of the proper time.  

In a second fit, the data were fitted leaving free 
$\dmd$, $\epsilon_d$, $R_1$ and $R_2$. 
The parameter $a$
was fixed to the value of $8.5\: 10^{-3} \rm  ps^{-1}$ obtained in the previous fit to the
simulation.  The results for the different parameters are: 
  $\epsilon_{d}=0.575 \pm 0.009$ (0.579), 
  $R_1 = 0.550 \pm 0.005 $  (0.554)  and 
  $R_2=0.080 \pm 0.022$  (0.059). Within parentheses are given the results  for the
 fit to the simulated data. 

\begin{figure}[htb]
\vspace*{-0.5cm}
\begin{center}
\epsfysize10.0cm\epsfxsize10.0cm
\epsffile{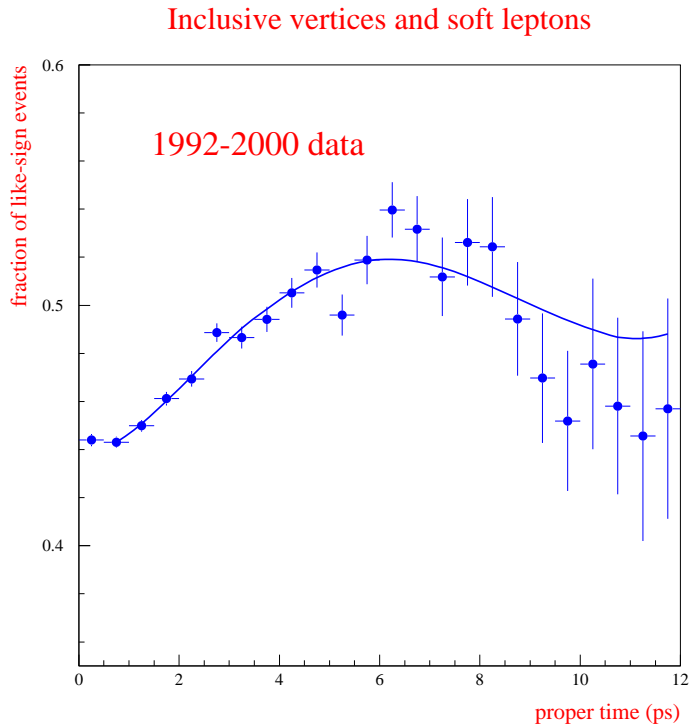}
\caption{Fraction of like-sign tagged events as a function of the reconstructed 
proper time using 
1992-2000 data. The data were shown as points with error bars, the solid line corresponds
to the fit.}
\label{dmdfit}
\end{center}
\end{figure}

The result for the $\mbox{B}_d$ mass difference is $\Delta m_d$ = 0.531 $\pm$ 
0.025 (stat.) 
 with a  $\chi^2/ ndf$ of 22.5/(23-4), as shown in Figure \ref{dmdfit}.

 The reason for performing a four parameter fit is that both tagging purities 
for 
 the $\mbox{B}^0_d$ meson and for the other B particles are determined using 
the data. Therefore systematic 
uncertainties on these parameters were largely reduced.
  In this way the fit results become also less sensitive to, for example, the fraction of
 $\mbox{B}^0_s$ mesons.
 Due to the fact that the fit was first applied and tuned to the simulated data, the resolution function is taken into account.

  A breakdown of the systematic errors affecting the measurement is given  
in Table \ref{sysdmd}. The range of values for the fractions and lifetimes of 
the different B species come from ref. \cite{ref:bosc}. 
 The fractions of $\mbox{B}^0_s$ mesons and B baryons were 
changed (correspondingly the other B fractions are recalculated) as
 well as 
 the lifetimes and backgrounds.
 The tagging correction factor $C$ (see section 2.6) was varied by a relative 5\% for the soft lepton sample and by 15\% for the inclusive vertex sample.
 The proper times were scaled with 1\%, coming from the scale uncertainty
on the reconstructed decay length and momentum, and the corresponding systematic
error on $\Delta m_d$ was -0.0049  ps$^{-1}$.    
 The resolution function $\cal{R}$ (see section \ref{ssec:fitting}) was smeared by an additional Gaussian term
 with a width of 0.1+0.03$t \rm \: ps$ and the resulting shift in $\Delta m_d$
was 0.0037  ps$^{-1}$.  This additional smearing corresponds to a 10\% systematic error on the expected decay length resolution and a 10\% systematic error on the expected momentum
 resolution. 
 
 The total systematic error amounts to  
 0.007  ps$^{-1}$.

 \begin{table}[htb]
 \begin{center}
 \vspace*{-1cm}
 \begin{tabular}{|c|c|c|} \hline
  error source & values & systematic error (ps$^{-1}$) \\ \hline 
  $f_{\rm B_s}$ & 0.097 to 0.108 \cite{ref:bosc}  &           -0.00021 \\
  $f_{\rm B\:baryon}$ &  0.103 to 0.12 \cite{ref:bosc} &       0.00039 \\
 $\tau_{\rm B_s} $ & 1.55 to 1.60 ps \cite{ref:bosc} & 0.0001 \\
 $\tau_{\rm B_u} $ & 1.65 to 1.67 ps \cite{ref:bosc} & -0.0008 \\
 $\tau_{\rm B_d} $ & 1.55 to 1.58 ps \cite{ref:bosc} & 0.0012 \\
 $\tau_{\rm B\:baryon}$ & 1.2 to 1.25 ps \cite{ref:bosc} & -0.0008 \\
 $uds$ background scale factor & 1 to 1.10  &       -0.00022 \\
 charm background scale factor & 1 to 1.10 &       0.00052\\
 tagging factor $\Delta C/C $ & variation 5\% (15\%)  &    0.0006 \\
 scale factor proper time & 1 to 1.01 &    -0.0049\\
 resolution smearing &  &   0.0037\\
\hline 
  Total systematic error  & & 0.0067 \\
 \hline
\end{tabular}
\caption{The systematic errors affecting the $\Delta m_d$ measurement.}
\label{sysdmd}
 \end{center}
 \end{table}
    
 The final result is thus:
 
\begin{center}

  $\Delta m_d$ = (0.531 $\pm$ 0.025 (stat) $\pm$ 0.007 (syst.)) ps$^{-1}$. 

\end{center}
 The total error is therefore 0.027 ps$^{-1}$.   

 This result for the 
 mass difference between the two physical states in the  
 $\mbox{B}^0_d-\overline{\mbox{B}^0_d}$ system   
 is compatible with those from other experiments \cite{ref:bosc}.

 A fit was also performed to extract the width difference 
 $\Delta \Gamma_{\rm B_d}$. 
 In the fit, the expression in Eq. (\ref{eq:ar1})
     $[ (1-2\epsilon_d) \cos(\Delta m_d t) + 1 ]$ was replaced 
 by      $[ (1-2\epsilon_d) \cos(\Delta m_d t) + \cosh(\Delta\Gamma_{\rm B_d} t /2) ]$
 and the expression in Eq. (\ref{eq:ar2}) was modified accordingly.
 The result of the five parameter fit is $|\Delta\Gamma_{\rm B_d}| = ( 0.00 \pm 0.06 ) \rm ps^{-1}$.  
 The total systematic error was evaluated for the error sources listed in
 Table \ref{sysdmd} and found to be $0.0002 \:\rm  ps^{-1}$. 
 Using the measured $\mbox{B}_d$ lifetime $\tau_{\rm B_d} = ( 1.55 \pm 0.03 )\rm  ps$ \cite{ref:bosc},  
 $|\Delta\Gamma_{\rm B_d}| / \Gamma_{\rm B_d} = 0.00 \pm 0.09 \:(tot)$.  
 The following upper limit was derived: 

\begin{center}
 $|\Delta\Gamma_{\rm B_d}| / \Gamma_{\rm B_d} < 0.18$ at 95\% CL. 
\end{center}

  \subsection{Search for $\mbox{B}^0_s-\overline{\mbox{B}^0_s}$  oscillations} 
  \label{ssec:searchbs}
\begin{figure}[htb]
\vspace*{-0.5cm}
\begin{center}
\epsfysize14.0cm\epsfxsize14.0cm
\epsffile{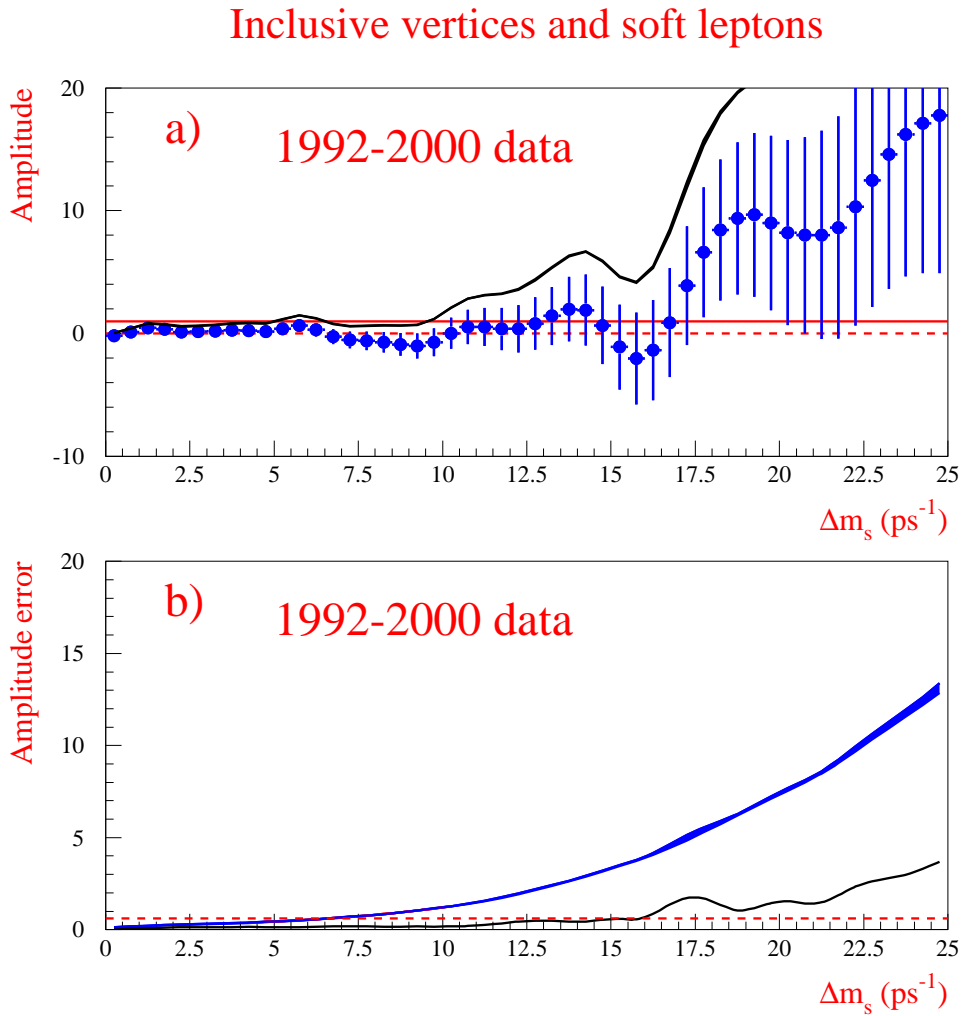}
\caption{a) Fitted values of the oscillation amplitude $A$ as a function of $\Delta m_s$. The horizontal line
corresponds to the value $A$=1. 
The black band is situated between the curves for $A+1.645\:\sigma_{A_{stat}}$ 
and $A+1.645\:\sigma_{A_{tot}}$. 
b) The total amplitude
error as a function of $\Delta m_s$. 
The upper band  is situated between the statistical error 
($\sigma_{A_{stat}}$) 
and the total error ($\sigma_{A_{tot}}$). 
The lower curve shows the 
systematic error $\sigma_{A_{sys}}$. 
The crossing point with the dashed line of the rising curve for the total error with $\sigma_{A}$=1/1.645 at 
$\Delta m_{s}=6.6 \:\rm  ps^{-1}$ gives the sensitivity.}
\label{amp1}
\end{center}
\end{figure}

\begin{figure}[htb]
\vspace*{-0.5cm}
\begin{center}
\epsfysize14.0cm\epsfxsize14.0cm
\epsffile{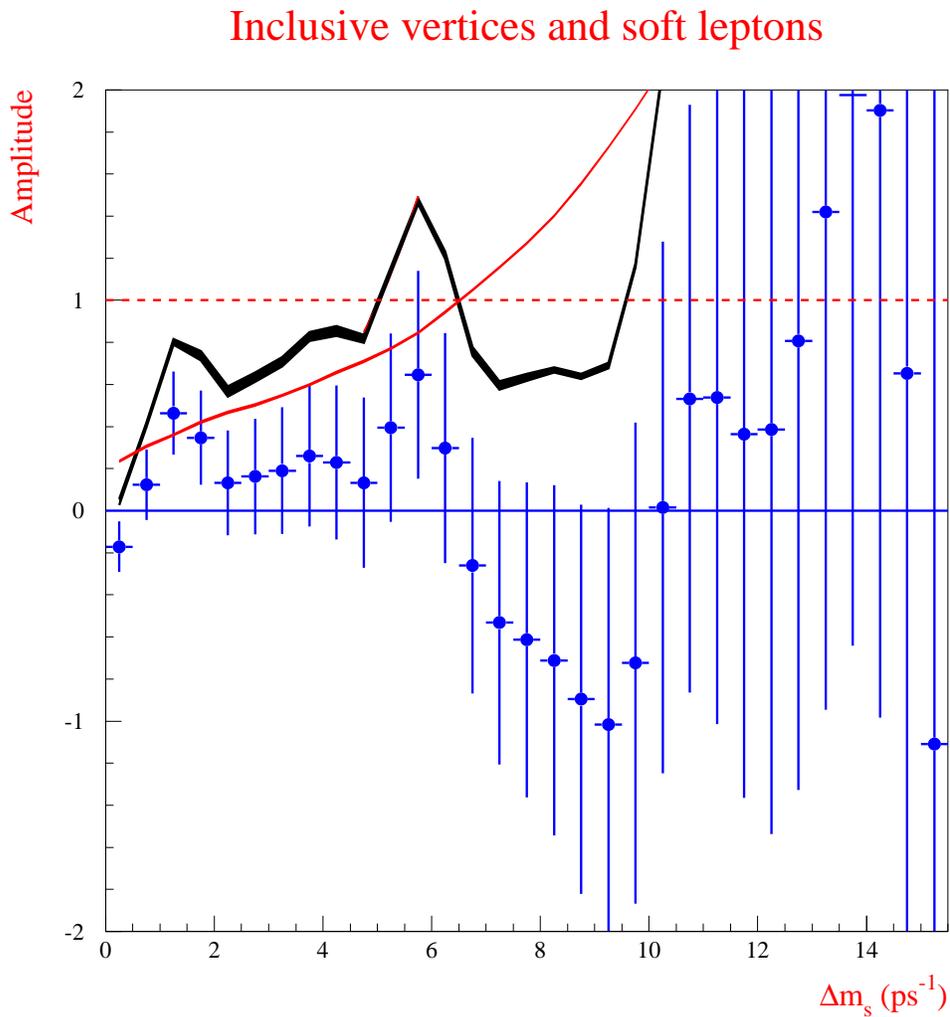}
\caption{Fitted values of the oscillation amplitude $A$ as a function of $\Delta m_s$. The data are identical to those of Fig.\ref{amp1}. 
The dashed horizontal line
corresponds to $A$=1. 
The black band is situated between the curves for $A+1.645 \:\sigma_{A_{stat}}$ 
and $A+1.645\: \sigma_{A_{tot}}$. 
The smoothly rising curve corresponds to  
$1.645 \:\sigma_{A_{tot}}$. 
The crossing point with $A$=1 at 
$\Delta m_{s}=6.6 \:\rm  ps^{-1}$ gives the expected lower limit at 95\% CL.}
\label{amp2}
\end{center}
\end{figure}

 To search for $\mbox{B}^0_s-\overline{\mbox{B}^0_s}$  oscillations a likelihood fit was performed,
 where the negative log-likelihood is defined as:

 \begin{equation} 
 {\cal L} = - \sum_{like-sign} ln({\cal P}^{like}(t_{rec},P_{comb},P_{decay}))  
  - \sum_{unlike-sign} ln({\cal P}^{unlike}(t_{rec},P_{prod},P_{decay})),  
 \end{equation}  
where the expression for ${\cal P}^{like}$ and ${\cal P}^{unlike}$ can
be found in Eqs. (\ref{likes}) and (\ref{unlikes}). 
 To extract results from this fit the so-called amplitude method 
 was used \cite{ref:amplitude}. For the mixed and unmixed $\mbox{B}_s^0$ events  the following
 expressions were used:

\begin{equation} 
{\cal P}_{\rm B_s^0}^{unmix.}~=
~\frac{1}{2 \tau_{\rm B_s}} e^{- \frac{t}{\tau_{\rm B_s}}} [ 1 + A \cos ({\Delta m_s t} ) ]
\end{equation}
and similarly:
\begin{equation} 
{\cal P}_{\rm B_s^0}^{mix.}~=
~\frac{1}{2 \tau_{\rm B_s}} e^{- \frac{t}{\tau_{\rm B_s}}} [ 1 - A \cos ({\Delta m_s t} ) ].
\end{equation}
 B$_s$ oscillations will correspond to a value A of unity. 
 The oscillation amplitude $A$ and its error $\sigma_A$ were fitted to the data as a function
 of $\dms$.    
 The result of the amplitude fit is shown in Figures \ref{amp1} and \ref{amp2}.

 Before discussing the result and its interpretation, the systematic errors 
 have been studied.
 This was done by changing one parameter at a time (for example $f_{\rm B_s}$) and 
redoing the full  amplitude fit.
 The systematic error was then evaluated as \cite{ref:amplitude}:
\begin{equation} 
 \sigma_A^{syst}= A_1-A_0+(1-A_0)\frac{\sigma_{A_{1}}^{stat} - \sigma_{A_0}^{stat}}{ \sigma_{A_{1}}^{stat}}, 
 \label{eq:asyst}
\end{equation}
 where $A_0$($A_1$) and $\sigma_{A_0}$($\sigma_{A_1}$) denote the fitted amplitude and error 
 before (after) changing the parameter. The last term in Eq.(\ref{eq:asyst}) takes into account the change in the error of the fitted amplitude. 
 The following parameters have been changed as in Table \ref{sysdmd}:\\
 \noindent 
 $\bullet$    $f_{\rm B_s}$ from 0.097 to 0.108,\\
 $\bullet$    the $uds$ and charm backgrounds have been scaled up by 10\%,\\
 $\bullet$    the tagging purity has been changed by varying the correction factor $C$ by 5\% for the soft
              lepton tag and by 15\% for the inclusive vertex tag, \\
 $\bullet$    the constant term $\sigma_0$ for $\cal{R}$, the width of the 
              resolution function (see section \ref{ssec:fitting}),  
              has been changed by a relative 10\%,\\ 
 $\bullet$    $\sigma_p$, the width of the resolution function for the 
              momentum, has been changed by a relative 10\%.

 The total systematic error as a function of $\dms$ is shown in 
Figure \ref{amp1}b. It is at most 35\% of the  statistical error.

 Using the results for the amplitude and its error it is
 possible to obtain the 95\% CL exclusion region or sensitivity. This region 
 corresponds to $A+1.645\: \sigma_A < 1$. This curve is shown in Figures \ref{amp1}
 and \ref{amp2}.
 No 
 $\mbox{B}^0_s-\overline{\mbox{B}^0_s}$  
oscillations have been
 observed in the data.
 A limit on the 
  mass difference of the two physical $\mbox{B}^0_s$ states can be put:
\begin{center}
 $\Delta m_s > 5.0 \:\rm  ps^{-1}$ at 95 \% CL. 
\end{center}
 Using the error on $A$, $\sigma_A$, the sensitivity is found to be: 
\begin{center}
 Sensitivity = 6.6 ps$^{-1}$.
\end{center} 
 The sensitivity would be 6.8 ps$^{-1}$ if the systematic error on
the amplitude was neglected\footnote{ 
In Table \ref{amptab} the results are given for the amplitude and its error as a function of $\dms$ after adjusting $f_{\rm B_s}$ to the recently published value of 0.106 \cite{PDG2002}.}.   

 \begin{table}[htb]
 \begin{center}
 \vspace*{-0cm}
 {\small 
 \begin{tabular}{|c|c|c|c|} \hline
   $\dms (ps^{-1})$  & $A$ & $\sigma_A$ (stat) & $\sigma_A$ (total)  \\ \hline
  0.25 &  -0.17 &   0.11 &   0.13 \\
  0.75 &   0.12 &   0.15 &   0.17 \\
  1.25 &   0.46 &   0.18 &   0.20 \\
  1.75 &   0.35 &   0.21 &   0.23 \\
  2.25 &   0.13 &   0.23 &   0.26 \\
  2.75 &   0.16 &   0.25 &   0.28 \\
  3.25 &   0.19 &   0.28 &   0.31 \\
  3.75 &   0.26 &   0.31 &   0.33 \\
  4.25 &   0.23 &   0.34 &   0.37 \\
  4.75 &   0.13 &   0.37 &   0.40 \\
  5.25 &   0.39 &   0.41 &   0.43 \\
  5.75 &   0.65 &   0.45 &   0.47 \\
  6.25 &   0.30 &   0.50 &   0.53 \\
  6.75 &  -0.26 &   0.56 &   0.58 \\
  7.25 &  -0.53 &   0.62 &   0.65 \\
  7.75 &  -0.61 &   0.69 &   0.71 \\
  8.25 &  -0.71 &   0.76 &   0.78 \\
  8.75 &  -0.90 &   0.85 &   0.87 \\
  9.25 &  -1.02 &   0.95 &   0.96 \\
  9.75 &  -0.72 &   1.05 &   1.06 \\
 10.25 &   0.02 &   1.16 &   1.17 \\
 10.75 &   0.53 &   1.28 &   1.30 \\
 11.25 &   0.54 &   1.43 &   1.45 \\
 11.75 &   0.36 &   1.59 &   1.62 \\
 12.25 &   0.39 &   1.77 &   1.82 \\
 12.75 &   0.81 &   1.96 &   2.02 \\
 13.25 &   1.42 &   2.17 &   2.23 \\
 13.75 &   1.98 &   2.40 &   2.44 \\
 14.25 &   1.90 &   2.65 &   2.69 \\
 14.75 &   0.65 &   2.91 &   2.96 \\
 15.25 &  -1.11 &   3.17 &   3.22 \\
 15.75 &  -2.04 &   3.44 &   3.48 \\
 16.25 &  -1.36 &   3.74 &   3.84 \\
 16.75 &   0.88 &   4.08 &   4.30 \\
 17.25 &   3.90 &   4.45 &   4.77 \\
 17.75 &   6.61 &   4.85 &   5.14 \\
 18.25 &   8.43 &   5.27 &   5.44 \\
 18.75 &   9.36 &   5.69 &   5.79 \\
 19.25 &   9.66 &   6.12 &   6.23 \\
 19.75 &   8.99 &   6.53 &   6.68 \\
 20.25 &   8.22 &   6.92 &   7.09 \\
 20.75 &   8.00 &   7.33 &   7.47 \\
 21.25 &   8.02 &   7.79 &   7.93 \\
 21.75 &   8.63 &   8.31 &   8.52 \\
 22.25 &  10.32 &   8.89 &   9.19 \\
 22.75 &  12.47 &   9.48 &   9.84 \\
 23.25 &  14.58 &  10.05 &  10.43 \\
 23.75 &  16.20 &  10.62 &  11.03 \\
 24.25 &  17.11 &  11.21 &  11.68 \\
 24.75 &  17.77 &  11.82 &  12.38 \\
 \hline
\end{tabular}
}
\caption{The amplitude and its statistical and systematic error as a function of $\dms$ after adjusting $f_{\rm B_s}$ to the published value of 0.106 \cite{PDG2002}.} 
\label{amptab}
\vspace*{-1cm}
 \end{center}
 \end{table}  

\clearpage 
 
\newpage
\section{A neural network analysis}

The inclusive $\mbox{B}^0_s$ analysis described in this section 
was an attempt to optimize the statistical precision attainable in the high 
$\Delta m_s$ region. 
This analysis made extensive use of neural network techniques for tagging and 
vertex reconstruction, mostly based on the BSAURUS \cite{BSAURUS} package. 
Several neural networks were used on the event and track level.
For optimal performance a good resolution on the proper time was required 
and this was achieved by keeping 
the energy and the vertex reconstruction separated in the analysis.
The separated treatment of decay length and energy reconstruction led to a 
CPU intensive two-dimensional integration for each event. 
Only the best class of events (in terms of the decay length resolution) was 
used, to reach an optimal performance for high $\Delta m_s$ values. 
The restrictive cuts on quality and decay length resolution led to a sample 
of only 30 k events for the data taken in 1994.

\subsection{Event selection} 
\label{sec:evsel}

Multihadronic Z$^0$ events were selected requiring at least 5 reconstructed 
tracks and a total reconstructed energy larger than 12\% of the centre-of-mass energy. 
The event was rejected if it had more than 3 jets or if 
the value of $|\cos (\theta_{thrust})|$ was larger than 0.75.
The cosine of the opening angle between the two most energetic jets was required
to be less than $-0.98$.  
Further, the value of the combined event b-tagging variable $x_{ev}$ as 
defined in ref. \cite{ref:borissov} had to be larger than $0.5$.
Events having an identified lepton with a transverse momentum larger than 1.2 
GeV/c were removed. 
To obtain a homogeneous data set, it was required that both the
liquid and gas radiators of the Barrel RICH were fully operational.

The same selection was applied to simulated Z $\rightarrow q\bar{q}$ events using the JETSET 7.3 \cite{ref:jetset} generator.

Each event was split into hemispheres using the plane perpendicular to the 
thrust axis. A first estimate of the  B candidate momentum vector was  
obtained by calculating the charged particles rapidities, 
with respect to the thrust axis, and by summing the momenta of those with rapidity $> 1.6$. 
In each hemisphere a secondary vertex was fitted using the tracks with vertex detector hits  from high rapidity
 charged particles. 
The secondary vertex fit was performed in three dimensions using as a constraint 
the direction of the B candidate momentum vector. 
The result of the vertex fit was used as an input to a Neural Network, the so-called TrackNet,   
that distinguishes between a fragmentation track and a track originating from a weakly decaying B hadron. 
In the final stage of the fit, the TrackNet output was used to add candidate tracks to the secondary vertex and the fit was redone. 

Finally, a hemisphere was rejected if   
the secondary vertex fit did not converge.

\subsection{Flavour tagging}
\label{tagging}

 The tagging of the quark flavour at production and decay times is necessary to distinguish mixed from unmixed $\mbox{B}^0_s$ mesons. 
Only the opposite hemisphere 
was used for the production tag to reduce correlations between the production
and decay tags. 
The decay tag was based on track-by-track flavour nets, which were later 
combined using a likelihood ratio to tag the presence of a B or $\bar{\mbox{\rm B}}$ meson at decay time in each hemisphere. For the production tag a dedicated neural network was used.

\subsection{The track-by-track flavour nets}
\label{subsec:track}

Eight different networks were trained corresponding to a production and a 
decay flavour network for each of the four B hadron types. The aim of each 
network was to exploit, track-by-track, the correlation between the charge 
of a single track and the b quark charge. 
This approach is motivated by the different decay chains of the various types 
of B hadrons where, for example, the `charge' of the $D$ meson determines 
the b quark charge. 

The discriminating input variables are: 
particle identification (e.g. kaon, proton and lepton probabilities), 
 B-D vertex separation based on a network trying to discriminate between 
tracks originating from the weakly decaying B hadron and those from the 
subsequent cascade $D$ meson decay, the momenta in the B rest frame and 
variables related to tracking quality. 
The track decay flavour nets use 21 input variables in total, while the track production flavour nets have 
18 input variables; essentially the same input variables without the lepton 
identification and the B-D net variables.

To obtain a flavour tag in a given hemisphere the individual track probabilities 
$P(track)^j_i$ ($i=\rm B_u,B_d,B_s,B_{baryon}$ and $j=$ production or decay) coming from the different networks were combined in the following way,

\begin{equation}
\label{eq:like}
 P(hem)^j_i = \sum_{tracks} q(track) \log \frac{1+P(track)^j_i}{1-P(track)^j_i},
\end{equation}
where $q(track)$ is the charge. For the production flavour tag, tracks 
with TrackNet values less than $0.5$ are selected, while for the decay flavour tag, tracks must have a TrackNet value above $0.5$. 

\subsection{The $\mbox{B}^0_s$ production and decay flavour tag}

The production flavour net was constructed using all the information available in the hemisphere,  i.e. 
the fragmentation and decay flavour probabilities $P(hem)^{prod,decay}_{\rm B_u,B_d,B_s,B_{baryon}}$,
and the quality of the information for the selected hemisphere.  
More details on the flavour networks and on the flavour tag can be found in \cite{BSAURUS}.

\begin{figure}[htb]
\begin{center}
  \epsfig{file=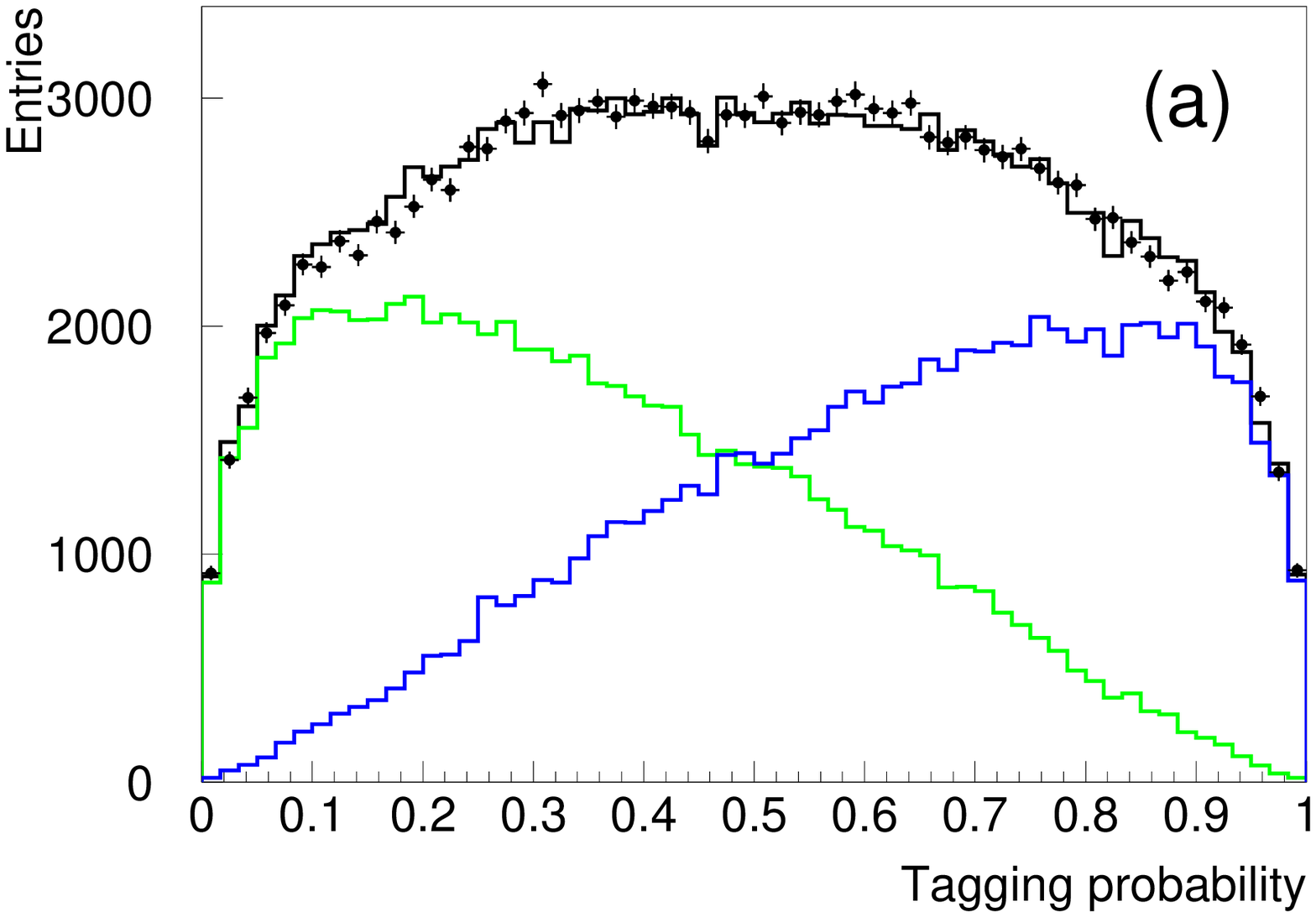,width=0.4\textwidth}
  \epsfig{file=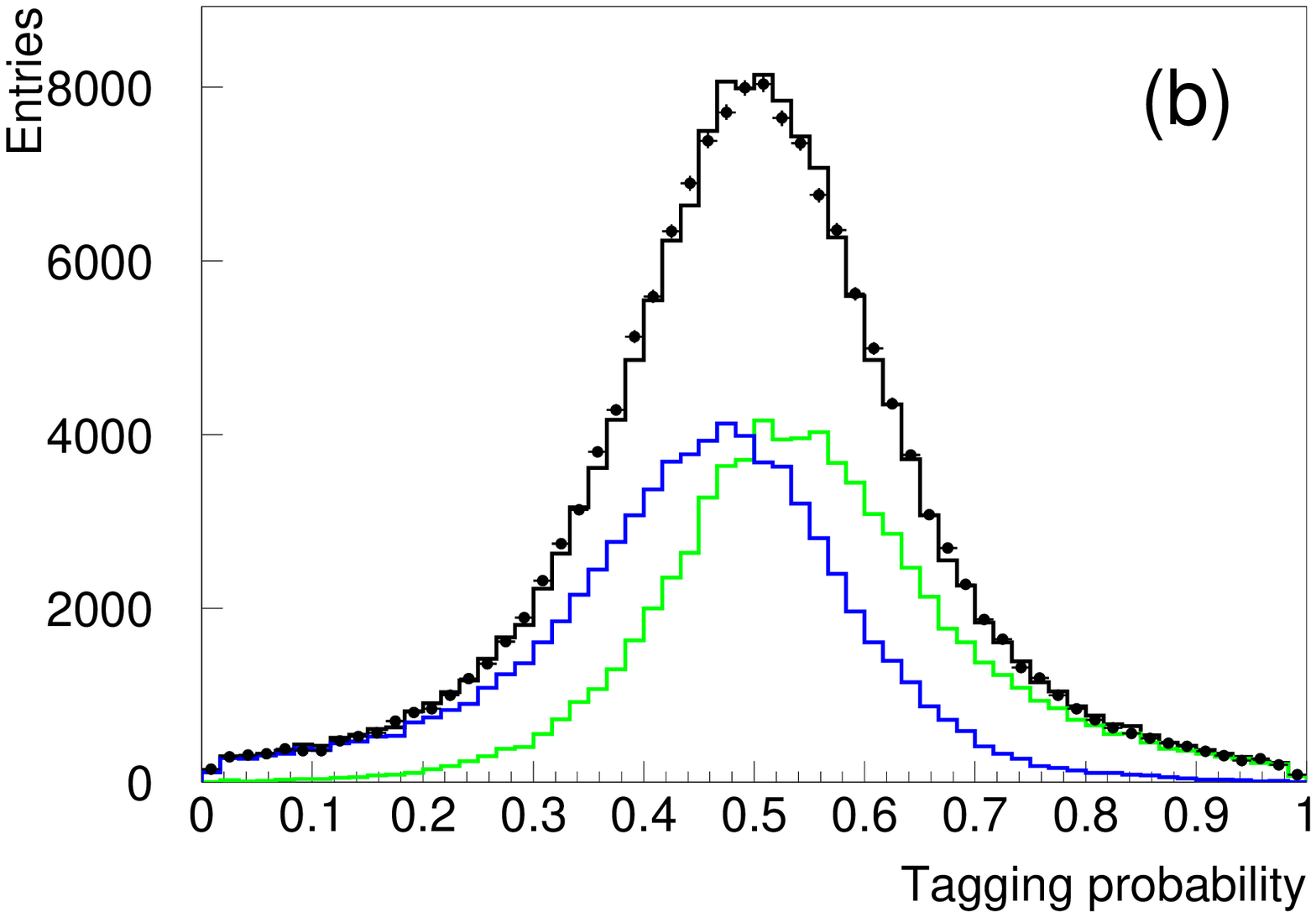,width=0.4\textwidth}
\caption[Production and decay tag distribution]{\label{fig:decay} The 
production (a) and decay (b) tag distributions for 1994 data and simulation.
The distributions for b and $\bar{\rm b}$ are indicated in light and dark grey. The decay tag
is optimized for ${\rm B_s}$ and $\overline{ \rm B_s}$ particles.}
\end{center}
\end{figure}

In Figure \ref{fig:decay}a the probability distribution for the production 
tag for 1994 data and simulation is shown. The grey lines indicate the distributions for b and $\bar{\rm b}$ quarks. The achieved tagging purity on simulation is 71\% at 100\% efficiency.

For the $\mbox{B}^0_s$ decay flavour tag, the probability $P(hem)^{decay}_{B^0_s}$ was used. 
In Figure \ref{fig:decay}b the $\mbox{B}^0_s$ decay flavour probability distributions for 1994 data and simulation are shown. 
The tagging purity on simulation was 62\% at 100\% efficiency.
The contributions from light and charm quarks are very small due to the high b purity of the sample of 98.3 \%.

\subsection{Energy reconstruction}
\label{subsec:erec}

To determine the proper time, a precise estimate of the energy of the 
 decaying B hadron is needed. 
The starting point was a raw estimate of the B energy $E_{raw}$ and mass 
$m_{raw}$. 
These quantities were determined by  weighting (with a sigmoid threshold function) the momentum and energy 
components of the charged particles by the TrackNet output value and the neutral particles 
by their rapidity. For three-jet events only the rapidity was used as a weight.
In this way particles coming from the decaying B hadron receive a higher
weight. 

The raw energy was corrected as a function of $m_{raw}$ and of the fraction of the 
energy in the hemisphere, $x_h$, to obtain an improved estimate of 
the energy.
This was done in the following way. The simulated data were divided 
into several samples according to the measured ratio $x_h$ and 
for each of these samples the $\Delta E$, defined as the true energy minus the raw energy, was plotted as function of $m_{raw}$. 
The median values of $\Delta E$ in each bin of $m_{raw}$ were calculated and 
the $m_{raw}$ dependence was fitted by a third order polynomial:
\begin{equation}
\Delta E(m_{raw},x_h)=a+b(m_{raw}-\left<m_{raw}\right>)+
c(m_{raw}-\left<m_{raw}\right>)^2+d(m_{raw}-\left<m_{raw}\right>)^3.
\end{equation}
The four parameters ${a,b,c,d}$ in each $x_h$ bin were then studied 
as functions of $x_h$ and parametrized with third and 
second-order polynomials. 
In this way a smooth correction function was obtained.

 This procedure led to an  estimate of the B hadron energy. 
Studies on simulated B events showed a large correlation between 
the number of tracks and the B energy resolution.
For this reason, the number of tracks in the hemisphere was chosen to define 
different resolution classes. 
In total 16 different classes were defined, starting with 2 tracks per 
hemisphere in class one and ending with 17 and more tracks in class 16. 
The central Gaussian of a double Gaussian fit to the B energy resolution 
varies from 4\% in the best class to 15\% in the worst. 

\subsection{Decay length reconstruction}
\label{subsec:decl}

Starting from the secondary vertex algorithm, described in section \ref{sec:evsel}, 
an optimized algorithm was developed with the aim of improving the decay 
length resolution and of minimizing the forward bias resulting from the 
inclusion of tracks from the cascade $D$ decay vertex in the B decay vertex 
reconstruction. 
Based on the output of the B-D net, a so-called `Stripping' algorithm was 
developed.

For the `Stripping' algorithm candidate tracks were selected if they had 
a TrackNet output larger than $0.5$ and a B-D net output value less than 
$0.45$. 
The B-D net cut value corresponds to an efficiency of 50\% for selecting a 
track 
from a weakly decaying B hadron at a purity of 75\%. 
A secondary vertex fit was performed if two or more tracks were selected. 
If the fit failed to converge within the algorithm criteria and more than two tracks were selected, the track with highest $\chi^2$ contribution was 
removed and the fit was repeated. 
This procedure was done iteratively until convergence was reached or two tracks were left. 
Finally, the direction of the B, as estimated by the B energy algorithm, 
was used as a constraint.  
The overall efficiency to find a vertex was about $50$\%.

Events with a very good decay length resolution were selected by requiring
that the expected error on the decay length was smaller than 200 $\mu m$.

Because of cuts on the TrackNet output, on the B-D output and on the expected 
decay length
error,  
less events will be reconstructed at small decay length. 
Therefore an acceptance function depending on the true B decay length 
was calculated using the simulation. 


After having applied these cuts, 30k hemispheres were selected in the 1994 
data sample. The b purity of the sample was estimated from simulated events 
to be 98.3\%.

\subsection{The likelihood fit}
\label{sec:lik}

In the fitting program, the like- and unlike-sign events were separated in the same way as described in section \ref{ssec:fitting} of the previous analysis and the same expressions for like- and unlike-sign probabilities were used.

A difference from the previous analysis was the treatment of the resolution 
functions ${\cal R}(l_{rec}-l_{true},l_{true})$ and 
${\cal R}((p_{rec}-p_{true})/p_{true})$, which were kept separated. 
As a parameterization for the decay length $l$ two asymmetric Gaussian distributions were 
chosen, while for the momentum reconstruction two symmetric Gaussian 
distributions were used. 
The probability for a B event to be observed at a proper time $P(t_{rec})$ 
is a convolution over an exponential B decay distribution, 
an acceptance function $A(l,p)$, the true B hadron momentum distribution $F(p)$ and the resolution functions ${\cal{R}}_{l}$ and ${\cal{R}}_{p}$, all four taken from simulation: 
\begin{equation}
{\cal P}_b (t_{rec}) = \int_{l=0}^{\infty} \int_{p=0}^{\infty} A(l,p) F(p) {\cal R}_{l}(l_{rec}-l,l){\cal R}_{p} ((p_{rec}-p)/p) \frac{e^{-l m/(\tau p)}}{\tau} dl dp,
\end{equation}
were $\tau$ denotes the B lifetime and Eq. (\ref{for:tau}) was used to calculate the proper time.

\subsection{Modelling simulation and data}

As explained in section \ref{Modelling} it is important to model
 precisely the tagging purities. 
In this analysis the raw purities were modified using 
a parameter $\alpha$ as defined in Eq. (\ref{eq:slope}). 
The decay and production tag parameters for the different particles were obtained 
from simulation, and are listed in Table \ref{tab:slope}. 
\begin{table}[t]
\begin{center}
\begin{tabular}{|c|c||c|c|}
\hline
particle & decay tag $\alpha_D$ & particle & production tag $\alpha_P$ \\
\hline
$\mbox{B}^0_s$ & 1 & b quarks & 0.94  \\
$\mbox{B}_d$   & 1.08 & c quarks & 0.56  \\
$\mbox{B}_u$   & 1.15 & uds quarks & 0.84  \\
$\mbox{B}_{baryon}$ & 0.93 & & \\
c quarks & 1.05 & & \\
uds quarks & 0.08 & & \\
\hline
\end{tabular}
\caption[Different quark type slopes]{\label{tab:slope} The $\alpha$ parameters for the decay and production tag for the different particles as obtained from the 1994 simulation} 
\end{center}
\end{table}

For the real data, the correction factor $C$, defined 
in Eq. (\ref{eq:slopedata}), was determined from the fraction of like-sign events, using the same method as was discussed in section \ref{Modelling}. 
Two correction factors were needed, one for $\mbox{B}_u$ mesons and one for 
the other B mesons. 
Their values were $C_{\rm B_u} = 0.53$ and $C = 0.81$.    

\begin{figure}[t]
\begin{center}
\begin{minipage}{0.47\textwidth}
  \epsfig{file=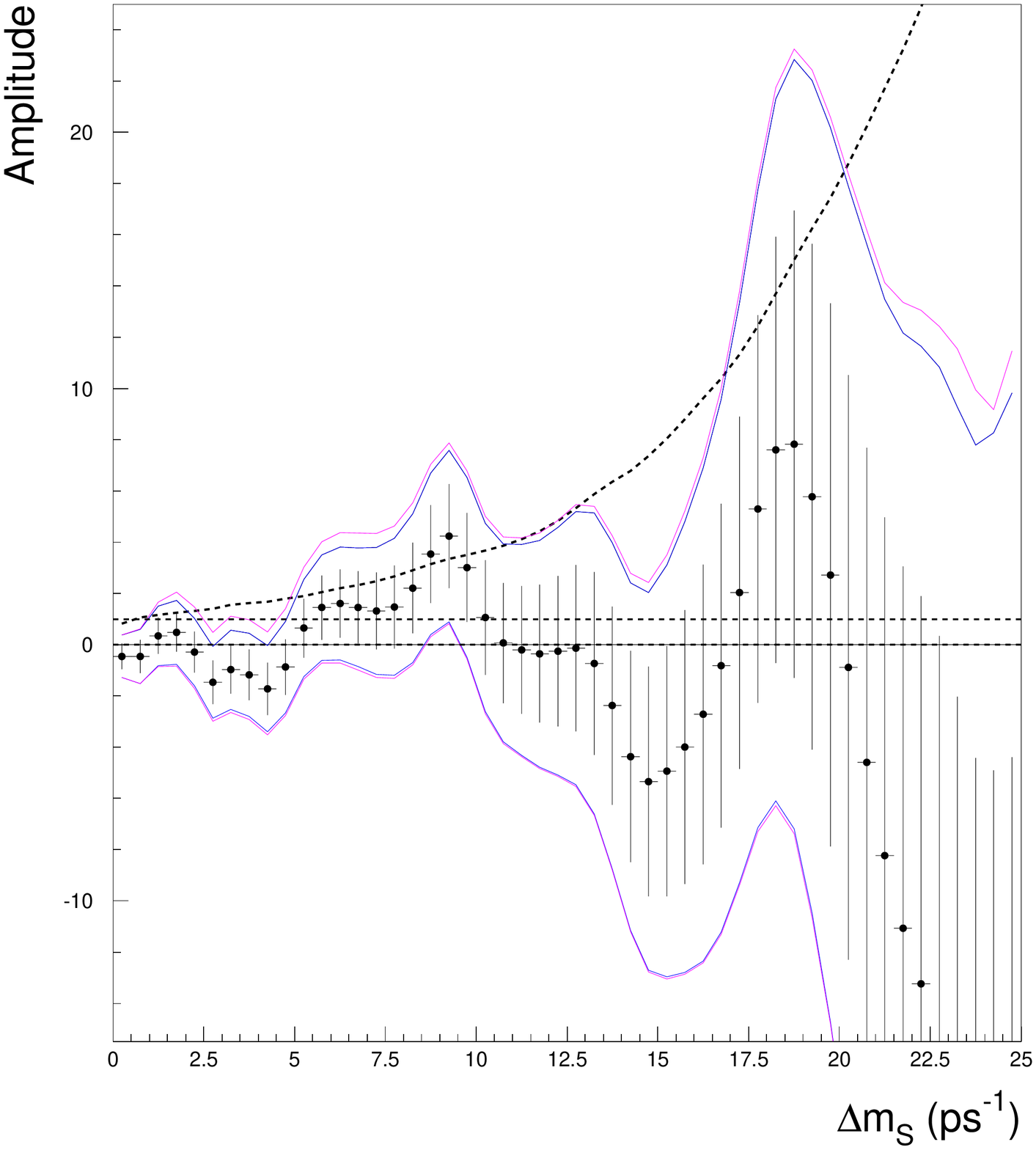,width=\textwidth}
\end{minipage}
\begin{minipage}{0.47\textwidth}
  \epsfig{file=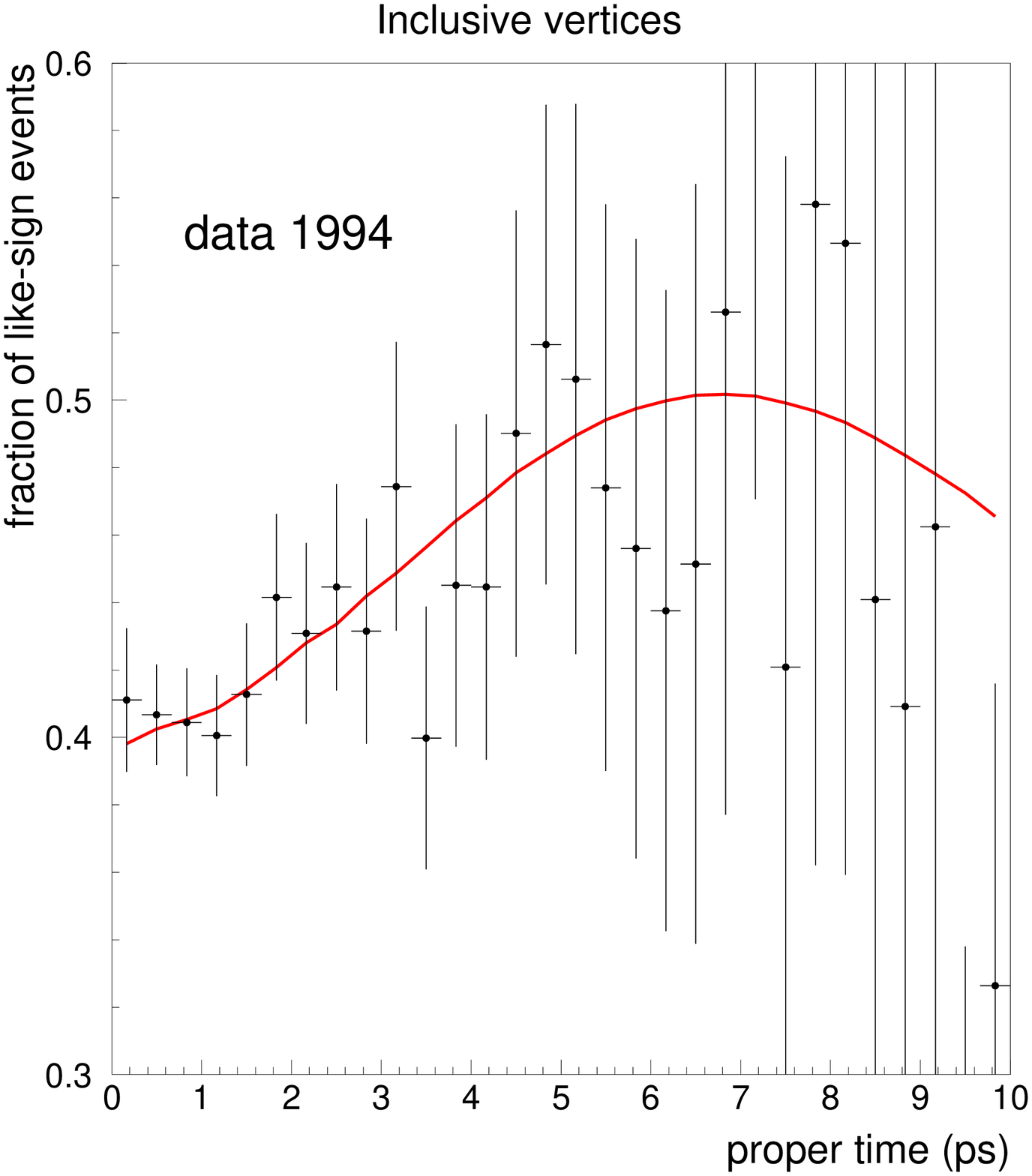,width=\textwidth}
\end{minipage}
\caption[Amplitude fit and like-sign ratio plot]{\label{fig:ampl} In the left plot the 
fitted $\mbox{B}^0_s$ oscillation amplitude for the NN analysis is shown  
as a function of 
$\Delta m_s$ as 
points and error bars. The continuous (dotted) lines correspond to $A \pm 
1.645 \sigma_{A_{stat (tot)}}$. The dashed line corresponds to $1.645 
\sigma_{A_{tot}}$. 
The plot on the right side shows the fraction of weighted like-sign tagged events as a function of the 
proper time. The data are shown as points with error bars, the parametrization 
is given as a solid line.}
\end{center}

\end{figure}

Using the amplitude method \cite{ref:amplitude} the result shown in Figure 
\ref{fig:ampl}a was obtained. A limit on $\dms$ was not extracted as the 
analysis was optimized for high values of $\dms$. 
Figure \ref{fig:ampl}b shows the agreement between the data and the 
description by the fitting programme.

Systematic uncertainties have been evaluated by varying a single parameter at a
time
(e.g $f_{\rm B_s}$) and redoing the full amplitude fit. The systematic error was 
then calculated as defined in Eq. (\ref{eq:asyst}).
The same parameters as described in section \ref{ssec:searchbs} were varied 
and the systematic error 
was determined to be at most 25\% of the statistical error.


The error on the fitted $\mbox{B}^0_s$ amplitude at $\dms$ of 15 and 20 ps$^{-1}$ gives respectively
5.1 and 11.8 for this analysis  
using only 1994 data. This can be compared with the
values of 
5.0 and 10.9 obtained with the previous analysis using only the 1994 data sample.  
The results of the neural network analysis optimized 
for high values of $\dms$ are compatible with 
the results for the 1992-2000 data 
shown in section \ref{ssec:searchbs}. No attempt is made to combine the results. 

\section{Conclusion}

 Using a total sample of 770 k events - of which 155 k events contain a soft lepton - the mass difference  
 between the two physical states in the $\mbox{B}^0_d-\overline{\mbox{B}^0_d}$
system 
was measured to be: 

\begin{center}
 $\Delta m_d = (0.531 \pm 0.025 (stat.) \pm 0.007 (syst.)) \rm  ps^{-1}$. 
\end{center}

The following limit on the width difference between the two states was obtained:
\begin{center}
 $|\Delta\Gamma_{\rm B_d}| / \Gamma_{\rm B_d} < 0.18$ at 95\% CL. 
\end{center}

As no evidence for  $\mbox{B}^0_s-\overline{\mbox{B}^0_s}$
oscillations was found, a limit on the 
mass difference of the two physical states was given:

\begin{center}
 $\Delta m_s > 5.0 \:\rm  ps^{-1}$ at 95 \% CL 
\end{center}
with a sensitivity equal to 6.6 ps$^{-1}$.

 These results are compatible with a neural network analysis
optimized for high values of $\dms$.  
 
\subsection*{Acknowledgements}
\vskip 3 mm
 We are greatly indebted to our technical 
collaborators, to the members of the CERN-SL Division for the excellent 
performance of the LEP collider, and to the funding agencies for their
support in building and operating the DELPHI detector.\\
We acknowledge in particular the support of \\
Austrian Federal Ministry of Education, Science and Culture,
GZ 616.364/2-III/2a/98, \\
FNRS--FWO, Flanders Institute to encourage scientific and technological 
research in the industry (IWT), Belgium,  \\
FINEP, CNPq, CAPES, FUJB and FAPERJ, Brazil, \\
Czech Ministry of Industry and Trade, GA CR 202/99/1362,\\
Commission of the European Communities (DG XII), \\
Direction des Sciences de la Mati$\grave{\mbox{\rm e}}$re, CEA, France, \\
Bundesministerium f$\ddot{\mbox{\rm u}}$r Bildung, Wissenschaft, Forschung 
und Technologie, Germany,\\
General Secretariat for Research and Technology, Greece, \\
National Science Foundation (NWO) and Foundation for Research on Matter (FOM),
The Netherlands, \\
Norwegian Research Council,  \\
State Committee for Scientific Research, Poland, SPUB-M/CERN/PO3/DZ296/2000,
SPUB-M/CERN/PO3/DZ297/2000, 2P03B 104 19 and 2P03B 69 23(2002-2004)\\
JNICT--Junta Nacional de Investiga\c{c}\~{a}o Cient\'{\i}fica 
e Tecnol$\acute{\mbox{\rm o}}$gica, Portugal, \\
Vedecka grantova agentura MS SR, Slovakia, Nr. 95/5195/134, \\
Ministry of Science and Technology of the Republic of Slovenia, \\
CICYT, Spain, AEN99-0950 and AEN99-0761,  \\
The Swedish Natural Science Research Council,      \\
Particle Physics and Astronomy Research Council, UK, \\
Department of Energy, USA, DE-FG02-01ER41155.



\newpage

\begin{thebibliography}{ref99}
\bibitem{ref:Franz} G. Altarelli and P.J. Franzini, Zeit. Phys {\bf C37} (1988) 271. \\
P.J. Franzini, Phys. Rep. {\bf 173} (1989) 1.
\bibitem{ref:bello} 
              M. Ciuchini et al., JHEP 0107:013, 2001 hep-ph/0012308. 
\bibitem{PDG2000} L. Wolfenstein, Phys. Rev Lett. {\bf 51} (1983) 1945. 
\bibitem{ref:bspapers} 
 DELPHI Coll., P. Abreu et al., Eur. Phys. J {\bf C18} (2000) 229, \\
 DELPHI Coll., P. Abreu et al., Eur. Phys. J. {\bf C16} (2000) 555, \\
 DELPHI Coll., P. Abreu et al.,  Phys. Lett. {\bf B414} (1997) 382.


\bibitem{ref:borissov} G. V. Borisov and C. Mariotti, Nucl. Instr. and Meth. {\bf
 A372}  (1996) 181,\\
G. V. Borisov {\sl Combined b-tagging}, DELPHI Note, PHYS 716-94 (1997).

\bibitem{ref:jetset} T. Sj\"{o}strand, {\sl PYTHIA 5.7 and JETSET 7.4}, Computer Physics 
 Commun. {\bf 82} (1994) 74. 

\bibitem{ref:delphi} DELPHI Coll., P. Aarnio et al., Nucl. Inst. Meth. 
{\bf A303} (1991) 233,\\
DELPHI Coll., P. Abreu et al., Nucl. Inst. Meth. {\bf A378} (1996) 57.


\bibitem{ref:tuning} DELPHI Coll., P. Abreu et al, Zeit. Phys. 
{\bf C71} (1996) 11. 

\bibitem{ref:detector} DELPHI Coll., ``DELSIM Reference Manual'', DELPHI 
87-97 PROG-100.

\bibitem{ref:bosc} ALEPH, CDF, DELPHI, L3, OPAL, SLD Coll.,
 ``Combined results on b-hadron production
         rates, lifetimes, oscillations and semileptonic decays'',
CERN-EP-2000-096.\\ 
 Recent updates http://lepbosc.web.cern.ch/LEPBOSC/
combined\underline{ }results/.

\bibitem{ref:amplitude} H. G. Moser and A. Roussarie, Nucl. Instr. and Meth. {\bf
 A384} (1997) 491.

\bibitem{BSAURUS} Z. Albrecht et al., {\sl The BSAURUS package}, hep-ex/0102001.
\bibitem{PDG2002} K. Hagiwara et al., Phys. Rev. {\bf D66}, 010001 (2002).  

\end{thebibliography}
\end{document}